\documentclass[twocolumn,floatfix,showpacs,preprintnumbers,amsmath,amssymb,article]{revtex4}

\usepackage{graphics}
\usepackage{graphicx}
\usepackage{dcolumn}
\usepackage{bm}
\usepackage{multirow}
\usepackage{amssymb}
\usepackage{perpage}	 

\MakePerPage{footnote}	 

\begin{document}

\title{New tools for the direct characterisation of FinFETs}
\author{G.C. Tettamanzi$^{1,2}$}
\email{g.tettamanzi@unsw.edu.au}
\author{A.\,Paul$^{3}$}
\author{S.\,Lee$^{3}$}
\author{G.\,Klimeck$^{3,4}$}
\author{S.\,Rogge$^{1,2}$}

\affiliation{$^{1}$Kavli Institute of Nanoscience, Delft University of Technology, Lorentzweg 1, 2628 CJ Delft, The Netherlands}
\affiliation{$^{2}$ CQC2T, University of New South Wales, Sydney, NSW 2052, Australia.}
\affiliation{$^{3}$Network for Computational Nanotechnology, Purdue University, West Lafayette, Indiana 47907, USA}
\affiliation{$^{4}$Jet Propulsion Laboratory, California Institute of Technology, Pasadena, California 91109, USA}

\begin{abstract}
\centering
This paper discusses how classical transport theories such as the thermionic emission (Ref.~\cite{sze1}), can be used as a powerful tool for the study and the understanding of the most complex  mechanisms of transport in Fin Field Effect Transistors (FinFETs). By means of simple current and differential conductance measurements, taken at different temperatures and different gate voltages ($V_G$'s), it is possible to extrapolate the evolution of important parameters such as the spatial region of transport and the height of thermionic barrier at the centre of the channel. Furthermore, if the measurements are used in conjunction with simulated data, it becomes possible to also extract the interface trap density of these objects. These are important results, also because these parameters are extracted directly on state-of-the-art devices and not in specially-designed test structures. The possible characterisation of the different regimes of transport that can arise in these ultra-scaled devices having a doped or an undoped channel are also discussed. Examples of these regimes are, full body inversion and weak body inversion. Specific cases demonstrating the strength of the thermionic tool are discussed in sections \ref{sec:II}, \ref{sec:III} and \ref{sec:IV}. This text has been designed as a comprehensive overview of 4 related publications \cite{Sel073502, Tet150,Tet2,Paul2011} and has been submitted as a book chapter in Ref.~\cite{Nadine}.
\end{abstract}

 \maketitle

\section{Transport in doped n-FinFETs} \label{sec:II}

Non-planar field-effect transistors called FinFETs \cite{His2320} have been developed to solve the issues of gate control encountered with the standard planar geometry when the channel length is reduced to a sub-45 nm size. Their triple-gate geometry is expected to have a more efficient gate action on the channel and to solve the leakage problem through the body of the transistor, one of the most dramatic short channel effects \cite{His2320}. However, their truly three-dimensional (3D) structure makes doping -and thus also potential- profiles very difficult to simulate and to understand using previous knowledge on device technology. Transport studies at low temperature, where the thermally activated transport is suppressed, can bring insight to these questions by measuring local gate action. For these reasons, in a recent work (Ref.~\cite{Sel073502}), the potential profile of these devices has been investigated by conductance measurements. This has allowed the observation of the formation of a sub-threshold channel at the edge of the silicon nanowire. This corner effect has been proposed \cite{Doy263,Fos745} as an additional contribution to the sub-threshold current in these 3D triple-gate structures, where the edges of the nanowire experience stronger gate action due to the geometric enhancement of the electric field. However, besides extensive simulation work \cite{Doy263,Fos745} -due to the difficulties with these 3D structures- very little experimental work \cite{Xio541} has been published previous to the ones discussed in this chapter. This paragraph focusses on the description of the experimental observation of the corner effect on doped devices identical to the ones described in Ref.~\cite{Sel073502} (see Fig.~\ref{fig:Chap2Part1Fig1} (a)).

\subsection{Thermionic emission in doped FinFET devices}

\begin{figure*}[htb]
\centering
\includegraphics[width=4.4 in,height=2.255 in]{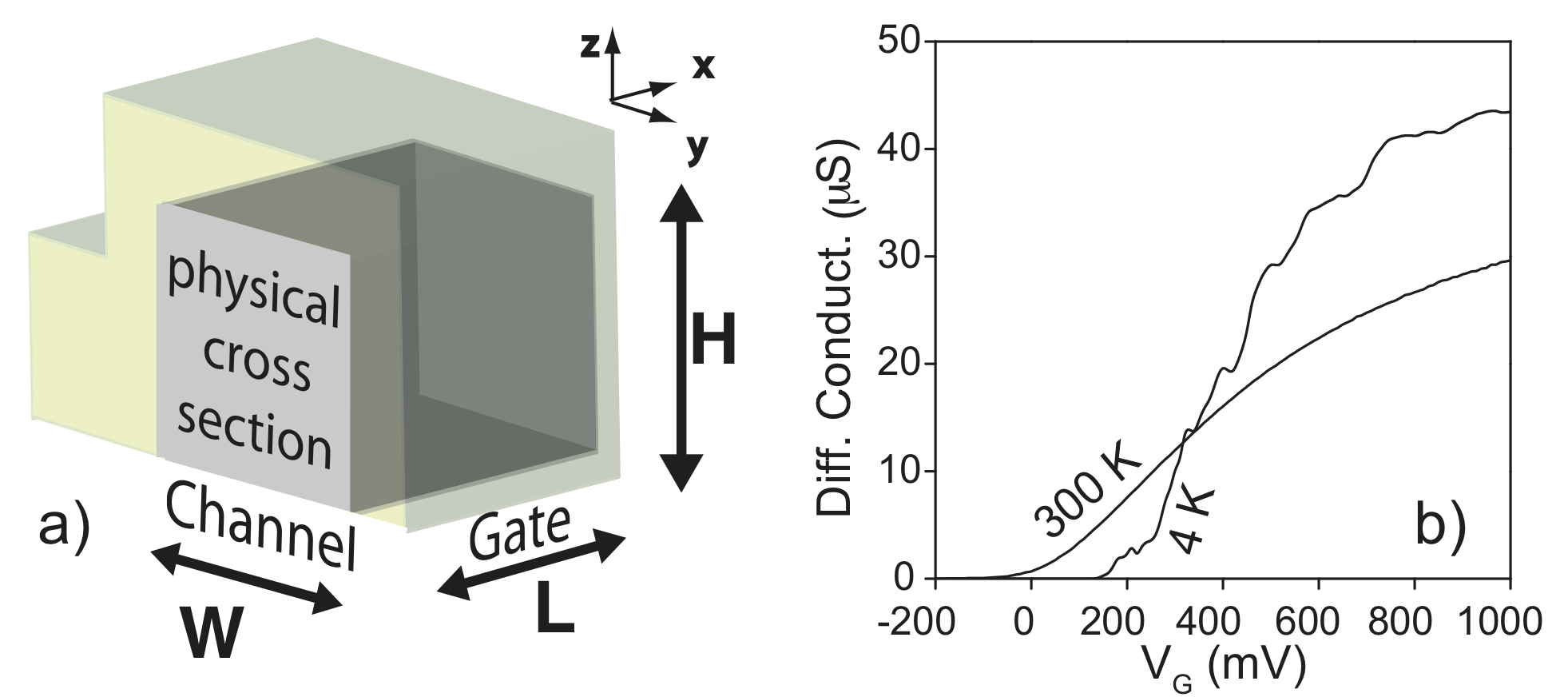}
\caption{a) Schematic of the FinFET geometry where the gate surrounds the Si nanowire (the fin). b) Low Bias differential conductance vs gate voltage for a long and narrow silicon FinFET ($L$ = 950 nm, $W$ = 35 nm)}
\label{fig:Chap2Part1Fig1}
\end{figure*}

The aim of this section is to show that, by using a combination of differential conductance ($G$=$dI_{SD}$/$dV_{SD}$) versus $V_G$ traces taken at different temperature, and of low temperature Coulomb blockade (CB) (see \cite{Sel073502} and references therein) measurements, it is possible to infer the existence of a dot located at the edge of the fin and thus of the corner effect \cite{Doy263,Fos745}. In the investigated device series the height of the fin wire is always $H$ = 65 nm, while the width ranges from $W$ = 35 nm to 1 $\mu$m and the gate length ranges from $L$ = 50 nm to 1 $\mu$m. The relatively high p-type doping ($\sim 10^{18}$ $cm^{-3}$) of the channel wire is chosen to ensure a depletion length shorter than half the channel length in order to have a fully developed potential barrier in this n-p-n structure and so to keep the conductance threshold at a large enough positive gate voltage. The characteristics at room temperature of these nanoscale FinFETs look therefore similar to those of their larger planar counterparts (see Fig. \ref{fig:Chap2Part1Fig1} (b) at 300 K). For sub-threshold voltages, a thermionic barrier ($E_b$) \cite{sze1} exists between the source and drain electron reservoirs and the transport is thermally activated at high enough temperature, as shown in Fig. \ref{fig:Chap2Part1Fig2} (a) and Fig. \ref{fig:Chap2Part1Fig2} (b). For very short devices, $G$ is simply given by the thermionic emission above the barrier according to the formula \cite{sze1}: 

\begin{equation}
\label{Gbulk}
 G_{3D} = S_{AA}A^{*}T\frac{e}{k_{B}}exp\Big(-\frac{E_{b}(V_{G})}{k_{B}T}\Big)
\end{equation}
\normalsize

\noindent where the effective Richardson constant A* for Si is 2.1$\times$ 120 A $cm^{2}$ $K^{2}$, $T$ is the temperature, $k_{B}$ the Boltzmann constant, e the elementary charge and $S_{AA}$ represents the active cross section, which can be interpreted as a good estimation of the portion of the physical cross section area through which transport preferentially occurs \cite{sze1}.

\subsection{Analysis of the thermionic regime (high temperatures)}

Several samples have been measured in this thermionic regime (80 K $\leqslant$ $T$ $\leqslant$ 250 K) and their conductance has been fitted using Eq.~(\ref{Gbulk}) to obtain $E_{b}$ and $S_{AA}$ (see Fig. \ref{fig:Chap2Part1Fig2} (c) and Fig. \ref{fig:Chap2Part1Fig2} (d)). The two 385 nm wide samples have the same cross section $S_{AA}$ $\approx$ 4 $nm^{2}$ although their length differ by a factor of 2. It is therefore possible to conclude that, in the sub-threshold regime transport is dominated by thermionic emission in these devices. The two 135 nm wide samples, however, have different $S_{AA}$ values, but this cannot imply a diffusive transport since the longest sample has the largest conductance. Another result is that the cross section $S_{AA}$ $\approx$ 4 $nm^{2}$ is much smaller than the channel width $W$ (135 or 385 nm) multiplied by the channel interface thickness (about 1 nm). This result is consistent with the corner effect that produces a lower conduction band (stronger electric field) along the two edges of the wire, where the current will flow preferentially (Fig. \ref{fig:Chap2Part1Fig3} (b)). The barrier height $E_b$ versus gate voltage is plotted in Fig. \ref{fig:Chap2Part1Fig2} (c). The data extrapolated to zero gate voltage are consistent with a 220 meV barrier height calculated for a p-type channel in contact with a n++ gate through a 1.4 nm $SiO_2$ dielectric \cite{sze1}. The linear dependence of the barrier height shows a good channel/gate coupling ratio, $\alpha$= $dE_{b}$/$(dV_G)$ = 0.68, due to the triple-gate geometry with a thin gate oxide. At higher gate voltage (above 300 mV), the coupling ratio decreases and a finite barrier survives up to large voltages. 

\subsection{Analysis of the Coulomb blockade regime (low temperatures)}

Analysis of the low-temperature transport (4 K $\leqslant$ $T$ $\leqslant$ 60 K, see Fig. \ref{fig:Chap2Part1Fig3} (c)) shows that the gate action remains constant inside the channel where localised states are formed. Two confining barriers are formed in the access regions (between channel and contacts), where the concentration of implanted arsenic atoms is reduced by the masking silicon nitride spacers placed next to the gate (see Fig. \ref{fig:Chap2Part1Fig3} (a)). For long channels and at low temperatures the conductance develops fluctuations versus gate voltage (see Fig. \ref{fig:Chap2Part1Fig1} (b)) with a pattern that reproduces after thermal cycling (at least for the main features). These fluctuations are caused by quantum interferences in the channel. For gate voltages close to the threshold, charge localisation occurs, especially for short fins. In fact, when short channel devices are cooled down to 4.2 K, conductance pattern develops a series of peaks, as can be seen in Fig. \ref{fig:Chap2Part1Fig3} (c), that can be attributed to Coulomb blockade of electrons in the potential well created in the channel by the two tunnel barriers of the low-doped access regions \cite{Sel073502}. This interpretation is supported by the channel-length dependence of the peak spacing discussed later. An explanation in terms of a quantum well formed by an impurity can be ruled out. An impurity or defect could not accept many electrons, i.e.: more than 20 for the 100 nm sample in Fig. \ref{fig:Chap2Part1Fig4} (b), since they represent a single charge or empty state.

\subsection{Interpretation of the results}

These results can be interpreted as follows; devices with shorter channel act as quantum dots where the conduction electrons are spatially localised and are Coulomb blockade for the transport by a finite charging energy bias. In the stability diagram of a quantum dot (see Fig. \ref{fig:Chap2Part1Fig3} (d)), the slopes of a triangular conducting sector give the ratios of the capacitances $C_{G}$, $C_S$, and $C_D$ between the dot and, respectively, the gate, source, and drain electrodes. In this way the dot/gate coupling $\alpha$= $C_G$ / ($C_{G}$ + $C_S$ + $C_D$) = 0.78 (0.65) for the first (second) resonance is found. These values are close to the channel/gate coupling of 0.68 obtained independently in 

\begin{figure*}[htb]
\centering
\includegraphics[width=4.45 in,height=4 in]{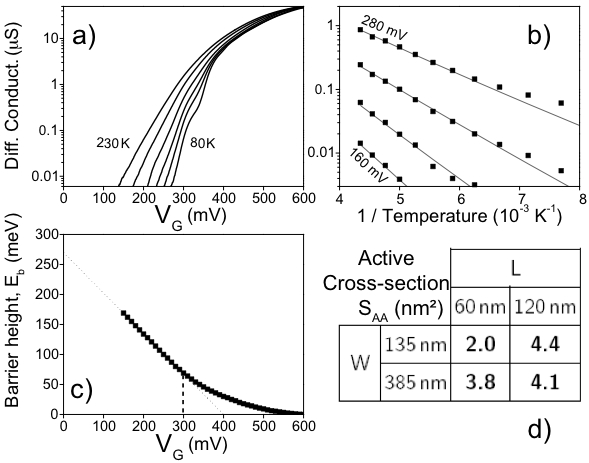}
\caption{a) Differential conductance vs gate voltage for a short and wide fin ($L$ = 60 nm, $W$ = 385 nm). b) Differential conductance plotted vs the inverse of the temperature for the same sample. The conductance is thermally activated above 150 K. c) Barrier height vs $V_G$ changing behaviour at 300 mV (same sample). d) Measured cross section $S_{AA}$ for the activated current of 4 samples with different lengths $L$ and widths $W$.}
\label{fig:Chap2Part1Fig2}
\end{figure*}

\noindent the same sample from the gate voltage dependence of the barrier height in the middle of the channel at higher temperatures. This result indicates that the gate coupling in the centre of the device remains constant and supports the idea of a minimum in the conduction band, as sketched in (Fig. \ref{fig:Chap2Part1Fig3} (b)). The peak spacing, $\Delta V_{G}$, is the change in gate voltage that increases by 1 the number of electrons in the dot located at the silicon/oxide interface. This quantity provides the dot/gate capacitance $C_G$ = e/$\Delta V_G$, and then the dot area $S$ = $C_G$/$C_{ox}$ using the gate capacitance per unit area $C_{ox}$ = $\epsilon _{ox}$/$t_{ox}$ = 0.025 F / $m^2$. The peak spacings for the same gate length ($L$ = 60 nm) but three different channel widths ($W$ = 35, 135, and 385 nm) can be compared in Fig. \ref{fig:Chap2Part1Fig4} (a). Although the patterns are not very regular, an average peak spacing of about 30 mV is obtained for all of them, indicating similar dot areas whereas the effective width is varied by more than a factor of 3.

\subsection{The corner effect}

The conductance patterns for three different lengths ($L$ = 60, 80, and 100 nm) shown in (Fig. \ref{fig:Chap2Part1Fig4} (b)) have decreasing average peak spacings ($\Delta V_G$ = 39, 24, and 6 mV, respectively) and therefore increasing dot areas ($S_{AA}$ = 160, 270, and 1100 $nm^2$). However, these areas are not strictly proportional to the gate length, so that the actual width could be length dependent or the actual dot length could be smaller than the gate length for very short fins. If it is assumed that the dot length equals the gate length, we obtain 2.7, 3.4, and 11 nm for the dot width, i.e., a small fraction of the total Si/oxide interface width $W_{eff}$ = $W$ + 2$H$ = 150 - 500 nm. The observation of similar dot widths of a few nanometers for different fin widths of hundreds of nanometers is consistent with the idea of a dot located at the edge of the fin and thus with the corner effect \cite{Doy263,Fos745}.

\subsection{Temperature dependence of the conductance peaks}

\begin{figure*}
\begin{center}
\resizebox{12 cm}{!}{\includegraphics{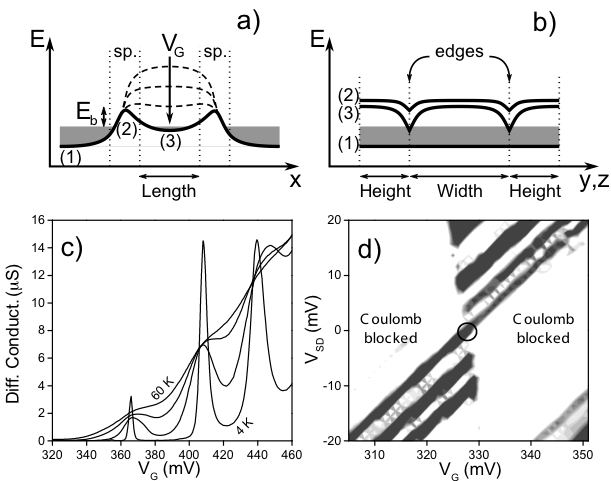}}
\caption{a) Conduction band edge profile with the highest barrier in the channel or in the access regions below the spacers (sp.) depending on the gate voltage. b) Band edge along the gate oxide interface (1) in the contacts, (2) in the barriers, and (3) in the channel. The corner effect produces two channels with low barriers at the wire edges. (c) Differential conductance vs gate voltage for a short and wide channel ($L$ = 60 nm, $W$ = 385 nm) showing Coulomb blockade peaks up to high temperatures (20 K steps). d) Stability diagram, i.e.: conductance vs gate and bias voltages at 4.2 K. The circle indicates a zero bias conductance peak, which develops into a triangular sector at finite bias.}
\label{fig:Chap2Part1Fig3}
\end{center}
\end{figure*}

\begin{figure*}[htb]
\centering
\includegraphics[width=4.25in,height=3.15 in]{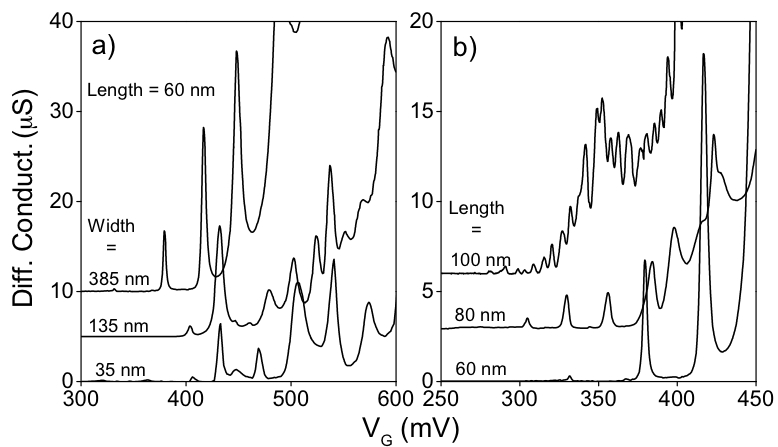}
\caption{Differential conductance vs $V_G$ at 4.2 K for several devices. a) Short fins ($L$= 60 nm) of different widths ($W$ = 35, 135, 385 nm) have a similar peak spacing. b) Devices with longer fins ($L$ = 60, 80, 100 $nm$) have a smaller peak spacing (the widths are different). The curves have been shifted for clarity.} 
\label{fig:Chap2Part1Fig4}
\end{figure*}

In addition to a large charging energy $E_c$ = $\alpha$ e$\Delta$ $V_G$, these dots also have a large quantum level spacing $\Delta E$, as can be deduced from the temperature dependence of the conductance peaks in Fig. \ref{fig:Chap2Part1Fig3} (c). When the temperature is lowered below the level spacing, the tunnelling process involves a single quantum level at a time and the peak height starts to increase above the high temperature value \cite{Sel073502}. The crossover from the classical to the quantum regime of Coulomb blockade being around 15

\noindent K, it is possible to estimate the level spacing to be about 1.3 meV. If the value $L$ = 60 nm is used for the gate length, in the expression $\Delta E$ = $3\pi ^2\hbar ^2$ / $2m^*L^2$ for the energy separation between the first and second states of a one-dimensional system, a level spacing $\Delta E$ = 1.6 meV, similar to the experimental estimation, is found.

This result supports the idea of a long dot extending over the whole gate length (assumed above to extract the dot width from the dot/gate capacitance).

\subsection{Conclusion of section II}

In doped channel FinFETs, experiments show the existence of a few nanometers wide edge channel, which shows itself in the activated current amplitude, the Coulomb blockade peaks spacing, and the quantum levels spacing. These channels are formed along the edges the devices due to an enhanced band bending called corner effect. To utilise the full FinFET cross section for electron transport with  a homogeneous current distribution, a lower sub-threshold current, and a larger on/off current ratio, this corner effect should be reduced. Better devices should have rounder corners on the scale of the depletion length and a lower doping concentration in the channel. 


\section{Transport in undoped n-FinFETS} \label{sec:III}

Section \ref{sec:II} showed that, in doped FinFET the geometry and the mechanisms of sub-threshold transport are affected by the presence of screening. This screening may results in a reduction of active transistor area (i.e.: corner effect) and in a sub-threshold swing (SS) degradation. Several models predicted that the introduction of an undoped channel FinFETs avoids the formation of the corner effect \cite{Doy263,Fos745} in these devices. However, we have found that even the undoped channel devices have a non-trivial, gate voltage ($V_{G}$) dependent current distribution, therefore there is a necessity to develop tools that could be used to investigate current distribution even in these intrinsic channel devices \cite{Tet150}. Design insights could be used to improve device characteristics towards their scaling to the nanometers size regime.

\subsection{Introduction to transport in undoped devices}

For undoped FinFETs and for widths smaller than 5 nm, full volume inversion is expected to arise (\cite{Won133} and references therein). Wider devices are expected to be in the regime of weak volume inversion (where the bands in

\noindent the channel closely follow the potential of $V_{G}$) only for $V_{G}$ $\ll$ $V_{th}$ \cite{Won133,Tau245}. Several groups have theoretically investigated the behaviour of such weak volume inversion devices using both classical \cite{Fos745}, and quantum \cite{Rui3369} computational models, but no experimental method that yields information on the location of the current-carrying regions of the channel exists prior to the work discussed in this section. Taur has studied this problem analytically for an undoped channel with double gate (DG) geometry, using a 1-D Poisson equation \cite{Tau245}. The main conclusion emerging from this work is that, when the gate voltage is increased, a crossover takes place between the behavior of the channel at $V_G$ $\ll$ $V_{th}$, and at $V_G$ $\sim$ $V_{th}$, caused by screening of induced carriers which subsequently increase the carrier density at the gate-channel interface. This section describes the first experimental observation of this prediction, furthermore the results of a 2D model are compared with experimental data, keeping in mind that the physical principles of this are fully analogues to the 1D case of Taur.

\subsection{Experimental results}

\begin{figure*}[htb]
\centering
\includegraphics[width=3.6 in,height=4.45 in]{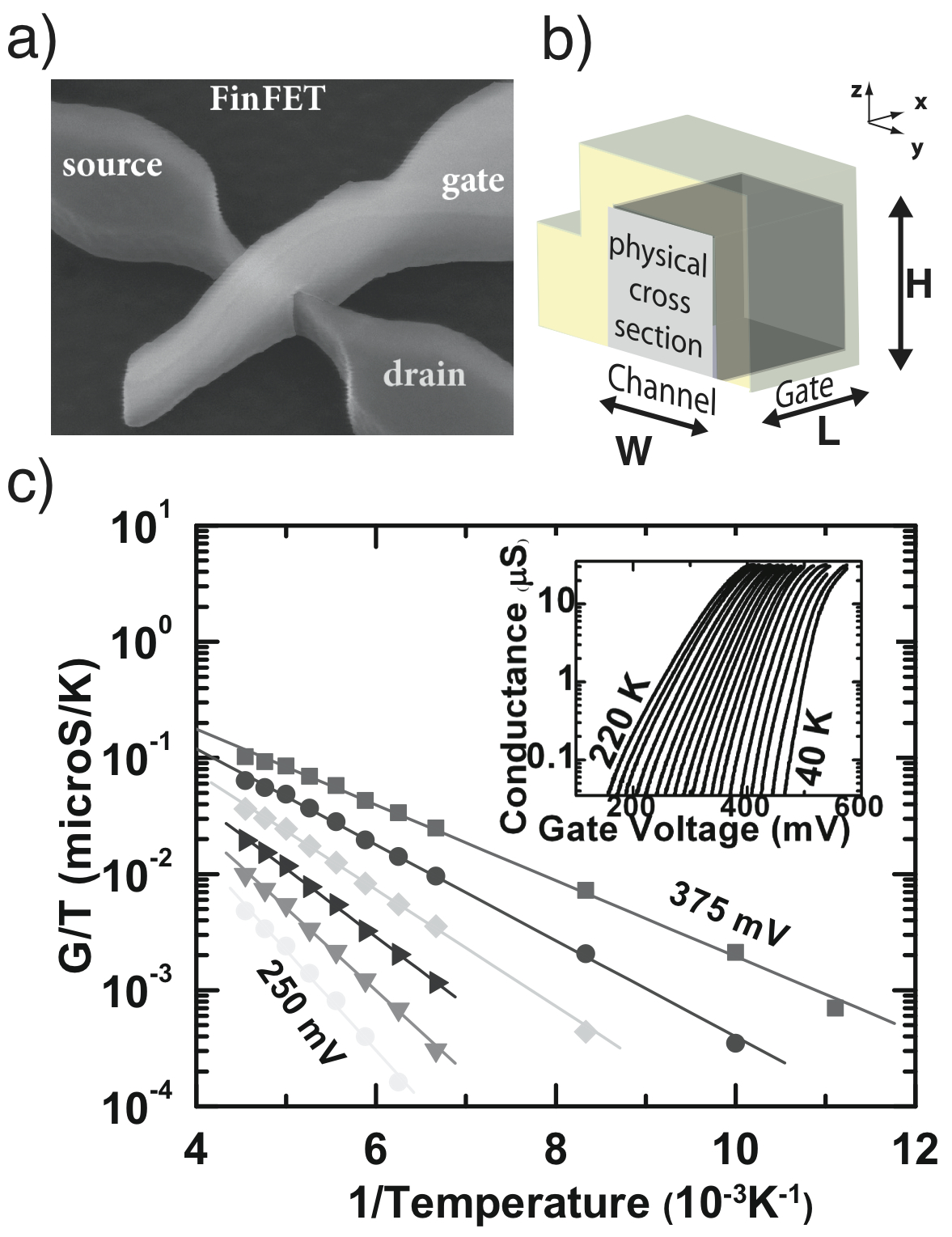}
\caption{a) Scanning Electron Microscope (SEM) image of typical FinFETs studied in this section. b) Schematic view of the FinFETs as in Fig.~\ref{fig:Chap2Part1Fig1}. The gate (light yellow) covers three faces of the channel (dark grey). L, H and W represent the channel length, height and width respectively. The physical cross-sectional area is shown in light grey. c) Fits used to extrapolate $E_b$ and $S_{AA}$ in one of our $W$ = 55 nm device. In the inset, differential conductance versus $V_{G}$ data, for different temperatures, are shown.} 
\label{fig:Chap2Part2Fig1}
\end{figure*}

Conductance versus temperature traces for a set of 8 undoped FinFET devices with the same channel length, ($L$ = 40 nm), and channel height, ($H$ = 65 nm), but different channel widths, ($W$ = 25 nm, 55 nm, 125 nm and 875 nm) are studied in this section. The discussion is focused on one device for each width since the same behavior for each of the devices of the same width is found consistently. The devices consist of a nanowire channel etched on a 65 nm Si intrinsic film with a wrap-around gate covering three faces of the channel (Fig. \ref{fig:Chap2Part2Fig1} (a) and Fig. \ref{fig:Chap2Part2Fig1} (b)) \cite{Col108}. They have a geometry identical to the ones of the previous section \ref{sec:II} \cite{Sel073502}, but their channels are completely undoped. In the devices of this study, an $HfSiO$ layer isolates a TiN layer from the intrinsic Si channel \cite{Col108}. Differential conductance data are taken at $V_{SD}$ = 0 mV using a lock-in technique. Fig. \ref{fig:Chap2Part2Fig1} (c) shows the $G/T$ versus 1000/$T$ data obtained from the $G$ versus $V_G$ data taken at different temperatures (inset in Fig. \ref{fig:Chap2Part2Fig1} (c)). Using the data of Fig. \ref{fig:Chap2Part2Fig1} (c), results for the source (drain)-channel barrier height, $E_{b}$, versus $V_{G}$ dependence and for the active cross-section area of the channel, $S_{AA}$, versus $V_G$ dependence can be extrapolated using the thermionic fitting procedure as described in section \ref{sec:II}. The important fact is that $S_{AA}$ can, also in the undoped case, be interpreted as a good estimation of the portion of the physical cross section area through which the transport preferentially occurs. Note that Eq. (\ref{Gbulk}) has only two parameters, $S_{AA}$ and $E_b$, and the accuracy obtained in the fits made using this equation \footnotemark[27] demonstrates the validity of the use of this model for the study of sub-threshold transport also in these undoped channel FinFETs. 

\footnotetext[27]{$R$ $\sim$ 0.99 for all fits of devices with width $\leqslant$ 125 nm, as shown in the Fig. \ref{fig:Chap2Part2Fig1} (c)}

\subsection{Evolution of the Barrier Height with Gate Voltage}

\begin{figure*}[htb]
\centering
\includegraphics[width=5.25 in,height=2.55 in]{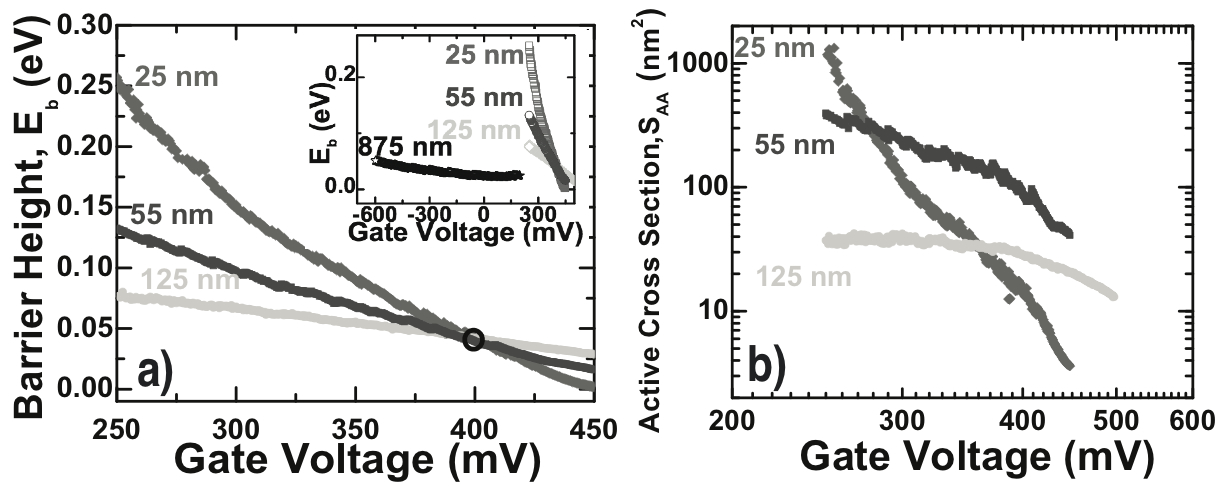}
\caption{Data obtained using the model of Eq. \ref{Gbulk}: a) $E_b$ versus $V_G$ for one device for each width from 25 nm to 125 nm. In the inset, calculated $E_b$ versus $V_G$ for all device widths are shown. b) Results of the dependence of the active cross section, $S_{AA}$, versus $V_G$ obtained for all devices with $W$ $\leqslant$ 125 nm.} 
\label{fig:Chap2Part2Fig2}
\end{figure*}

Fig. \ref{fig:Chap2Part2Fig2} (a) examines the barrier height as a function of $V_G$. An expected \noindent decrease in $E_b$ while increasing $V_G$ is observed (as for doped devices, see Fig. \ref{fig:Chap2Part1Fig2} (c)). The inset of Fig. \ref{fig:Chap2Part2Fig2} (a) shows that, this effect is less pronounced for a wider device. The decrease is to be attributed to short-channel effects (SCE's) that influence the electronic characteristics even at low bias. This trend is also reflected by the data of Table \ref{table0}, where the coupling factors obtained from our thermionic fits, $\alpha _1$=d$E_b$/d$V_G$ \footnotemark[28], show a decrease for increasing width. 

\footnotetext[28]{see also previous section \ref{sec:II}, thus the electrostatic coupling between the gate and the bulk of the channel}

\subsubsection{Capacitive coupling}

\noindent In Table \ref{table0}, the coupling between the potential of the channel interface and $V_G$, $\alpha _2$, extracted from Coulomb blockade (CB) measurements (at 4.2 K) of confined states that are present at the Channel/Gate interface \cite{Hin4637} is also shown. $\alpha _2$, is found to be a constant independent of $W$. In CB theory, $\alpha _2$ is the ratio between the electrochemical potential of the confined states and the change in $V_G$. This ratio can be estimated from the so called Òstability diagramÓ \cite{Sel073502} as it is shown in the previous section \ref{sec:II}. Overall, these results lead to the conclusion that the coupling to the channel interface remains constant for increasing $W$, whereas the coupling to the centre of the channel does not. In the 875 nm devices, SCEÕs are so strong (see inset Fig. \ref{fig:Chap2Part2Fig2} (a)), that the thermionic theory loses accuracy; hence the results of these devices will not be discussed any further. All the $E_b$ versus $V_G$ curves, as depicted in Figure \ref{fig:Chap2Part2Fig2} (a), cross each other at around 0.4 V (outlined by the black circle), before complete inversion of the channel takes place at $V_{th}$ $\sim$ 0.5 V \cite{Col108}. This suggests that for these devices and at $V_G$ = 0.4 V, the work function of the $TiN$ is equal to the affinity of the Si channel in our devices (flat bands condition). The same value has also been verified in other measurements using capacitance-voltage ($C$-$V$) techniques \cite{Sin332}, independently from the $W$ of the channel. This fact confirm, that, also for these devices, similarly to the ones described in section \ref{sec:II}, activated transport over the channel barrier is indeed observed. However, for these undoped devices, the barrier is formed by the Metal/Oxide/Semiconductor interface, which at $V_G$ = 0.4 V will not dependent on $W$. The crossing point in Fig. \ref{fig:Chap2Part2Fig2} (a) is not located exactly at $E_b$ = 0 meV, but is at 50 meV. This feature is attributed to the presence, at the Channel-Gate boundary, of interface states (already found in CB measurements) that can store charge, repel electrons and therefore raise-up the barrier by a small amount. In $Si/SiO^2$ systems that have been studied in the past, these states were estimated to give an energy shift quantifiable between 70 and 120 meV \cite{Hin4637}, in line with the data of this section.

\begin{table}[hbt]
\centering
\begin{tabular}{|c|l|c|l|}\hline
 & &    \\
 \textbf{Width (nm)}&$\alpha _{1}$&$\alpha _{2}$ \\
 & &    \\
\hline
25	& 1  & 0.7\\
\hline
55	& 0.7 & 0.8\\
 \hline  
125	& 0.14 & 0.8\\
 \hline   
875	& 0.03 & 0.8\\
 \hline  
\end{tabular}
\caption{Summary of the characteristics gate channel capacitive coupling of devices reported in this study, obtained from the results of Fit as in Fig 2a ($\alpha _{1}$) and from Coulomb Blockage (CB) measurements at 4.2 K  ($\alpha _{2}$).}
\label{table0}
\end{table}

\subsection{Evolution of the Active Cross Section with Gate Voltage}

The data of $S_{AA}$ for these undoped FinFETs show a surprising different evolution with increasing $V_{G}$'s if compared to what has been observed in the previous section \ref{sec:II} for doped channel devices. Fig. \ref{fig:Chap2Part2Fig2} (b) shows $S_{AA}$ as a function of $V_{G}$ extrapolated using Eq.~(\ref{Gbulk}). These results are then compared to the analytical model \cite{Tau245} discussed before and to the self-consistent simulations performed as described in \cite{Neo1286,Paul1,Lee1,Kli2090}. At low $V_G$, devices with $W$ = 25 nm show an active cross-sectional area of around 1000 $nm^2$ (see Fig. \ref{fig:Chap2Part2Fig2} (b)). This is almost equal to the physical cross sectional area of the channel at these widths. At higher $V_G$, the active cross-sectional decreases to a few $nm^2$. The interpretation of this data is as follows: at low $V_G$, transport in these devices is uniformly distributed everywhere in the physical cross-section of the channel (weak volume inversion). But with the increase of $V_G$, an increase of carrier density in the region near the interface, which leads to a reduction of $S_{AA}$, arise. This interpretation corresponds with the screening mechanism discussed in Ref. \cite{Tau245}. Subsequently the action of the gate on the centre of the channel is suppressed. Devices that have 55 nm and 125 nm widths behave in a fashion similar to the ones with 25 nm, but show a less pronounced decreasing trend and counter intuitive small values for $S_{AA}$, as a progressive reduction of $\alpha _1$ (i.e.: of the gate-to-channel coupling) for increasing $W$ is indeed observed. This is not a surprise as the barrier in these larger devices is lower and more carriers are allowed to migrate to the interface enhancing the screening effect. These results give, for the first time, an experimental insight into the mechanisms of conduction in undoped FinFETs. 

\begin{figure*}[htb]
\centering
\includegraphics[width=5.35 in,height=4.57 in]{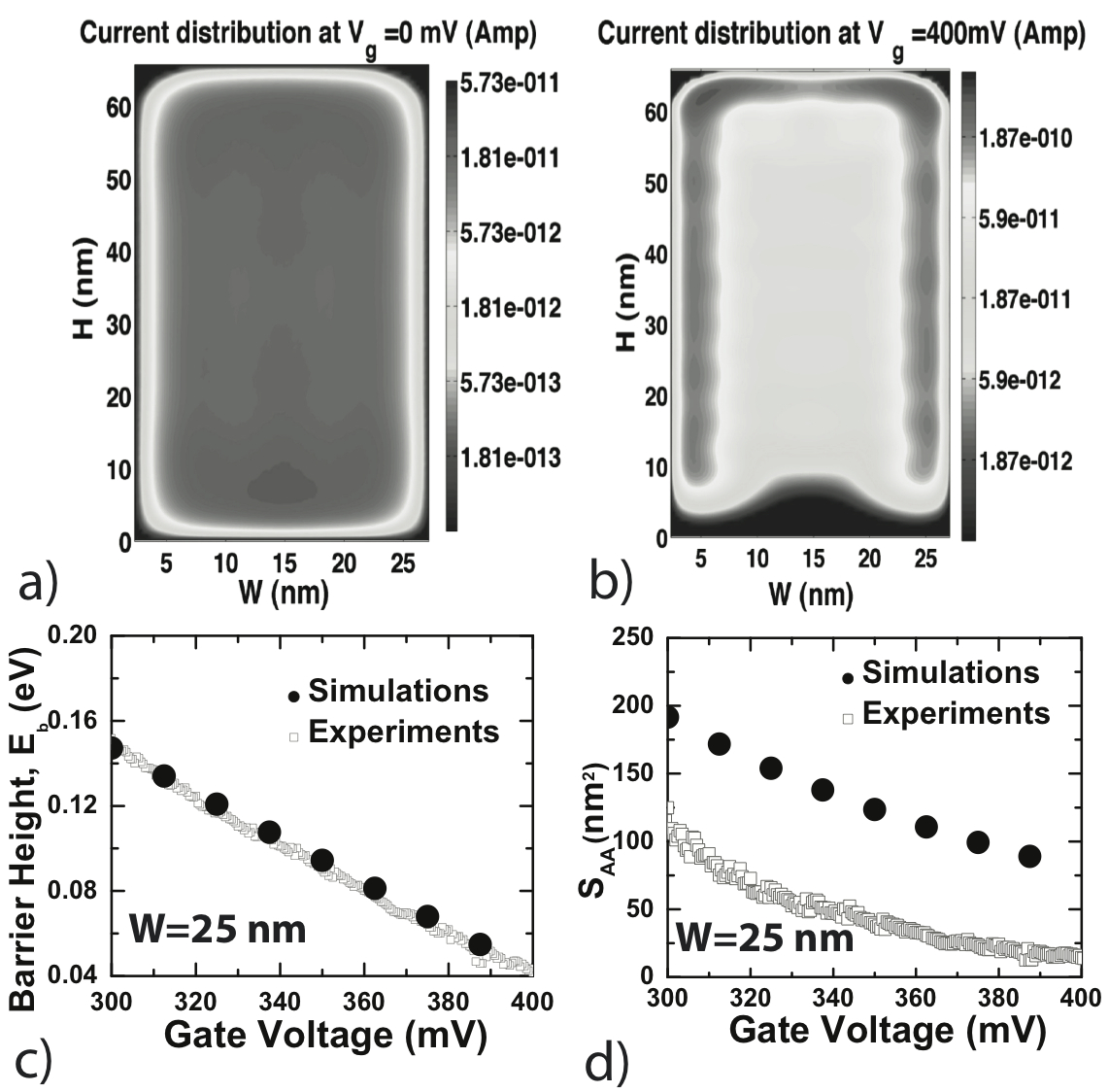}
\caption{Current distributions, for a) $V_G$ = 0 mV, b) $V_G$ = 400 mV, obtained using TB simulations for a geometry having $L$ = 65 nm and $W$ = 25 nm. Comparison of the simulated c) $E_b$ and d) $S_{AA}$ with the experimental data for a $W$ = 25 nm device.} 
\label{fig:Chap2Part2Fig3}
\end{figure*}

\subsection{Comparison with simulation}

State-of-the-art-simulations, done using an atomistic 10 band $sp3d5s^*$ Tight-Binding (TB) model \cite{Kli601,Lee1}, have been used to perform electronic structure calculation, coupled self-consistently with a 2D Poisson solver \cite{Neo1286}, and terminal characteristics using a ballistic top of the barrier (ToB) model \cite{Paul1} have been obtained. Due to the extensively large cross-section of the device that combines up to 44,192 atoms in the simulation domain, a new NEMO 3D code \cite{Lee1} has been integrated into the top of the barrier analysis \cite{Paul1}. This expanded modelling capability has made possible to compare experiment and simulations results. The effects of the variation of the potential in the source-drain direction are not expected to play a role in the simulated devices since $V_{SD}$ is very small \cite{Paul1,Tet150}. Also, the gate length is long enough to suppress the tunnelling current from source to drain \cite{Paul1,Tet150}. In fact, using a geometry identical to the one of the FinFETs used in these experiments, with $W$ = 25 nm, $H$ = 65 nm and under similar biases, the simulated current distribution shows a crossover from a situation of weak volume inversion at $V_G$ = 0 mV (Fig. \ref{fig:Chap2Part2Fig3} (a)) to a situation of transport confined prevalently at the interface at $V_G$ = 400 mV (Fig. \ref{fig:Chap2Part2Fig3} (b)). 

The simulated spatial current distribution (Fig.~\ref{fig:Chap2Part2Fig3}) gives a good indication of where the mobile charges predominately flow in the channel. From calculation too, a reduction of $S_{AA}$ with increasing $V_G$ is obtained, see Fig.~\ref{fig:Chap2Part2Fig3} (d). However, this reduction is not as sharp as in the experimental data, as these simulations have been performed at $T$ = 300 K and also due to the absence of interface states (expected to enhance the effect of screening in real devices as it will be discussed in the following section \ref{sec:IV}) \cite{Hin4637,Sel073502}. As a final benchmark to this experimental method, the results of the TB simulations have been used to calculate the current and the conductance at different temperatures and to extract, using again Eq.~(\ref{Gbulk}), simulated $E_b$ and $S_{AA}$ for a $W$ = 25 nm device. In fact, in Figure \ref{fig:Chap2Part2Fig3} (c) and \ref{fig:Chap2Part2Fig3} (d), the simulated values are compared to the experiments and it is found that it is possible to predict experimental results with good accuracy, although the simulations overestimate the values of $S_{AA}$ (probably for the same reasons discussed for Fig. \ref{fig:Chap2Part2Fig3} (b)). In any case, the comparison between experimental and simulation give a demonstration of the reliability of the method developed in this section \cite{Tet150}. This opens the way of its systematic use to obtain information about the magnitude and the position of carriers in FET devices in general and not only in FinFET structures. In these investigations, possible modifications of $A^*$ due to the constrained geometry \cite{Rag1261} of the devices have been neglected, as it is found to be negligible, and tunnelling regimes of transport \cite{App048301} have been excluded due to different temperatures dependences.

\subsection{Conclusion of section III}

In conclusion, the results presented in section \ref{sec:III} are the first experimental study of the behaviour of the active cross-section area as a function of $V_G$ for undoped FinFETs. In particular, conductance traces for a set of undoped FinFETs having the same channel length and height but different width, together with TB simulations for the device of $W$ = 25 nm have been presented. For all these small devices ($W$ $\leqslant$ 125 nm), a mechanism of inversion of the bands from flat band to band bending in the interface regions respectively, all as a function of $V_G$, has been proposed and demonstrated. Therefore this section discusses the first ever direct observation of the theoretical results suggested by Taur. The validity of thermionic approach as a tool for the investigation of sub-threshold transport in undoped FET devices has been confirmed and some answers to the fundamental technological questions, such as how to localise and quantify areas of transport have been provided.

\section{Interface trap density metrology of undoped n-FinFETs} \label{sec:IV}

\subsection{Introduction}

\begin{figure}[htb]
\centering
\includegraphics[width=3in,height=3in]{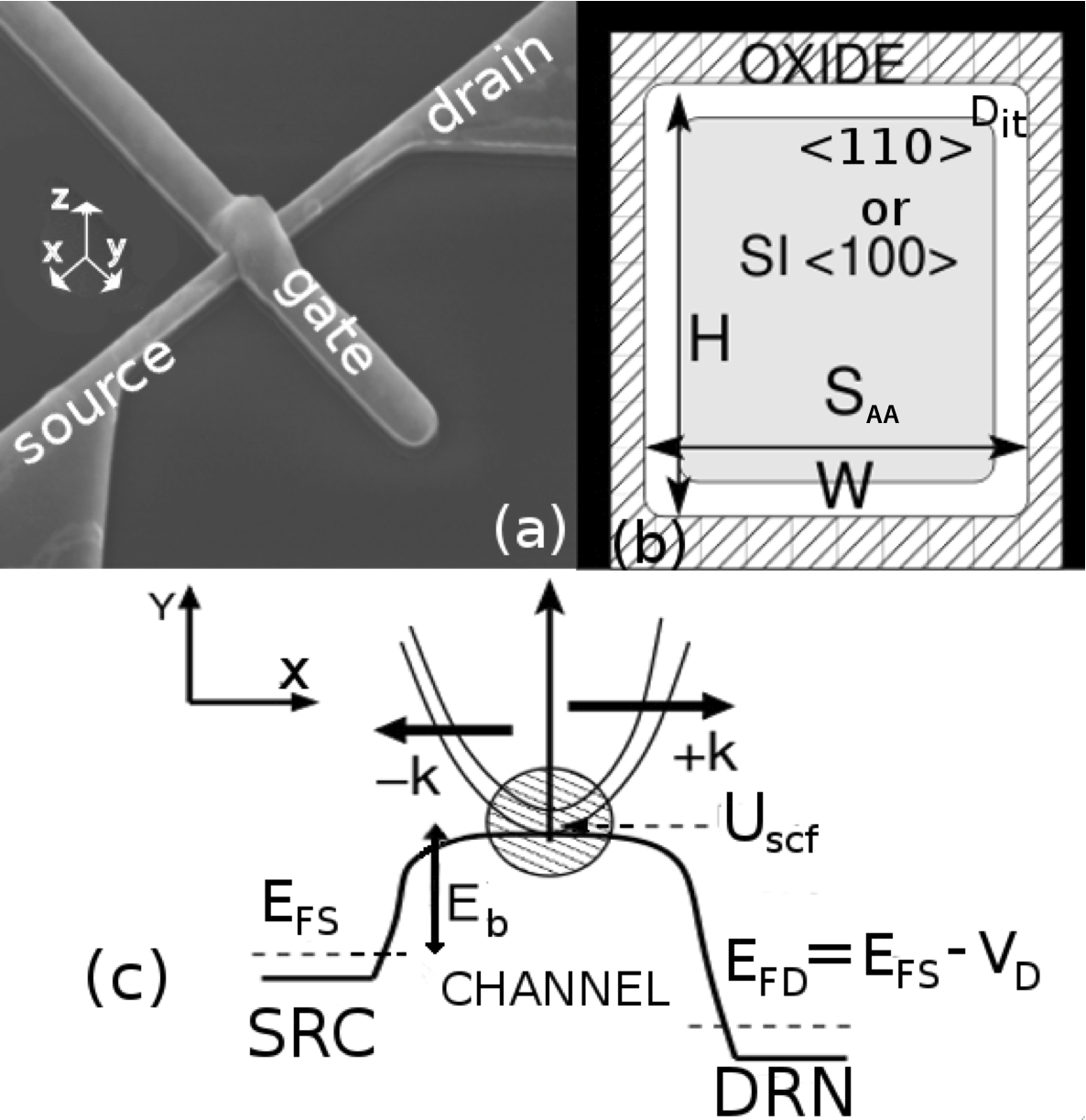}
\caption{(a) Scanning-electron-microscope (SEM) image of a Si n-FinFET with [100] channel orientation and single fin. (b) The schematic of the cross-sectional cut in the Y-Z plane of a typical tri-gated FinFET. The active cross-section ($S_{AA}$) is in gray, $H$ and $W$ are the physical height and width, respectively. (c) Ballistic top of the barrier model employed for calculating the thermionic current in the FinFETs.}
\label{fig:Finfet_cartoon}
\end{figure}

In the previous sections \ref{sec:II} and \ref{sec:III}, it has been demonstrated that, by using thermionic emission, it is possible to measure (1) the active channel cross-section area ($S_{AA}$) (see Fig. \ref{fig:Finfet_cartoon} (b)), and (2) the source to channel barrier height ($E_{b}$), hence opening new ways to investigate FinFETs. Furthermore, in section \ref{sec:III}, it was found that for undoped FinFETs, although the trends of the $S_{AA}$ values obtained by mean of experiments and of theoretical simulations were identical, differences in the absolute values were observed. These differences were found to be caused by the presence of interface states at the metal-oxide-semiconductor interface of the experimental devices \cite{Tet150,Lee186}. These states can trap electrons and enhance screening, therefore reducing the action of the gate on the channel, and as a final result, a decrease in the absolute value of $S_{AA}$ in the experimental data is observed. Typical $D_{it}$ frequency or time dependent measurements cannot be performed on ultimate devices but only on custom designed structures \cite{Kap232}. Such custom structures may only be partially reflective towards the possibly surface orientation-dependent and geometry-dependent $D_{it}$.

\subsection{Aim of the section}

In this section, a simple set of methods for the direct estimation of $D_{it}$ in ultimate devices is provided. The comparison between the values of $D_{it}$ obtained with these two methods and the values obtained using a method implemented in the past \cite{Kap232} show similar trends. A new approach to trap density metrology is of critical importance as CMOS scaling takes device dimensions into the nanometer regime. At these scales, quantities such as $D_{it}$ can vary rapidly with device geometry, rendering old techniques inadequate as they cannot be applied directly in these ultra-scaled devices.

\noindent In this section it is shown that, by using simple mathematical manipulations and the difference between experimental and simulated values of $S_{AA}$ and of the capacitive coupling, $\alpha$ (see previous sections), it is possible to infer the value of the interface trap density ($D_{it}$). Furthermore, to shed more light into the complicated transport phenomena that can arise in these undoped FinFETs, the work of previous sections is expanded and more careful investigations of the evolution of $S_{AA}$ and of $E_{b}$ are performed. For theoretically investigating these devices, the atomic representation is used. The band structure is obtained using a 10 band $sp^3d^5s^*$ Tight-Binding (TB) model with spin orbit coupling (SO) \cite{Kli601,Boy115201,Neo1286}, which is confirmed to be well suited for modelling the band-structure of these confined silicon channels, since TB can easily take into account the material, geometrical, strain and potential fluctuations at the atomic scale \cite{Neo1286}. This model takes also into account the coupling of the conduction and the valence bands which is neglected in simple models like the  effective mass approximation (EMA). As shown in Section \ref{sec:III}, semi-classical `Top of the barrier' (ToB) model accurately captures the thermionic transport  (Fig.~\ref{fig:Finfet_cartoon} (c)) \cite{Neo1286,Paul1,Tet150}, the same model can also shed more light on the inner details of the transport which is discussed next.


\subsection{New implementation of interface trap metrology}  

In the undoped devices studied here, qualitatively similar theoretical and experimental trends for the active cross section area versus $V_{G}$ and barrier height versus $V_{G}$ are found \cite{Tet150}. However, the theoretically obtained values quantitatively over-estimated the experimental values. The reduced experimental values can be attributed to the presence of interface traps in these FinFETs \cite{Tet2,Lee186,Kap232}. The effect of interface traps on the channel property are even more dominant in the extremely thin FinFETs \cite{Tet2}. In this section it is shown how this difference in $S_{AA}$ and $E_{b}$ can be utilised for the direct estimation of the interface trap density ($D_{it}$) in FinFETs, thereby eliminating the need to implement special FinFETs geometries to determine $D_{it}$ \cite{Kap232} and providing a new tool for performing interface trap metrology. 

This paragraph has been divided into the following sections. Section \ref{sec_dev} provides the details about the FinFETs for which interface trap density metrology has been implemented and the fundamentals of the experimental procedures which are in line with sections \ref{sec:II} and \ref{sec:III}. The details about the self-consistent calculations are provided in Sec. \ref{modeling_sec} and more insight on the theoretical extraction of $E_{b}$ and $S_{AA}$ is outlined in Sec. \ref{sec_mod_Eb}. Section \ref{extract_sec} provides the details of the two procedures for obtaining the interface trap density. The theoretical and experimental results and the discussion on them are given in Sec. \ref{sec_result}, while Sec. \ref{sec_current_res} discussed current distributions. The conclusions are summarised in Sec. \ref{sec_conc}.

\begin{table}[hbt]
\centering
\begin{tabular}{|c|l|l|l|c|l|}\hline
Label & H& W  & L & Channel & $H_{2}$  \\
      &(nm)   &(nm) & (nm)      & Orientation (X)& anneal\\\hline
A  & 65 & 25 & 40 & [100] & Yes \\\hline
B & 65 & 25 & 40 & [100] & No \\\hline
C & 65 & $\sim$5 & 40 & [100] & No \\\hline
D  & 40 & 18 & 40 & [110] & Yes \\\hline
E & 40 & 18 & 40 & [110] & Yes \\\hline
F & 40 & $\sim$3-5 & 40 & [110] & Yes \\\hline
G & 65 & $\sim$7 & 40 & [100] & Yes \\\hline
\end{tabular}
\caption{Table 2. Si n-FinFETs used in the trap metrology study along with their labels. The surface hydrogen annealing detail is also shown. The channel is intrinsic Si, while the source and the drain are n-type doped for all the FinFETs.}
\label{table1}
\end{table}

\subsection{Device and experimental details} \label{sec_dev}

The undoped n-FinFETs used in this work ($A-G$, see table \ref{table1}) consist of nanowire channels etched on a Si intrinsic film with a wrap-around gate covering the three faces of the channels (Fig. \ref{fig:Finfet_cartoon} (a)) \cite{Col108} identical to the ones discussed in section \ref{sec:III}. FinFETs with two different channel orientations of $[100]$ ((FinFETs A-C and G)) and $[110]$ ((FinFETs D-F)) have been used (see Table \ref{table1}). All the FinFETs have the same channel length ($L$ = 40 nm). The channel height ($H$) is either 40 nm or 65 nm (Table \ref{table1}). The channel width ($W$) varies between 3 to 25 nm. An HfSiO (high-$\kappa$) layer isolates a TiN layer from the intrinsic Si channel \cite{Col108}. These FinFETs have either one channel (FinFETs A-C and G) or ten channels (FinFETs D-F). These devices have two different surface treatments (with or without $H_{2}$ annealing) as shown in Table \ref{table1}.

\noindent \textit{Measurement procedure:} The experimental value of $E_{b}$ and $S_{AA}$ are obtained using the differential conductance method introduced in sections \ref{sec:II} and \ref{sec:III}. The conductance data are taken at $V_{SD}=0$ V using a lock-in technique. The full experimental method and the required ambient conditions have been outlined in detail in Ref.~\cite{Tet150}. In the next section we discuss the theoretical approach to calculate the values of $E_{b}$ and $S_{AA}$ in tri-gated n-FinFETs.

\subsection{Modelling approach}
\label{modeling_sec}

To obtain the self-consistent charge and potential, and transport characteristics in the n-FinFETs, the electronic structure is calculated using an atomistic 10 band $sp^3d^5s^*$ semi-empirical Tight-Binding (TB) \cite{Kli601} as discussed in the previous section \ref{sec:III}. Using thermionic fitting procedure \cite{Tet150}, $E_{b}$, $\alpha$ and $S_{AA}$ can be extracted using the experimental and theoretical conductance ($G$) using Eq.~(\ref{Gbulk}) for a 3D system \cite{sze1}. This equation will hold only when the cross-section size of the FinFET is large enough (i.e.: $W$, $H$ $>$ 20 nm) to be considered a 3D bulk system. In this study, $S_{AA}$ is extracted for FinFETs with W(H) $\approx$ 25 nm (65 nm). When the $3D$ approximation is not true anymore (i.e.: $W$ or $H$ $\lesssim$ 20 nm), only $E_{b}$ and $\alpha$ can be correctly extrapolated \cite{Tet150}. Since the FinFETs studied here show (i) negligible source-to-drain tunnelling current  and (ii) reduced SCEs \cite{Tet150}, the ToB model is applicable to such devices \cite{Paul1}. For the simulations, all the FinFETs are n-type doped in the source and drain to a value of 5$\times10^{19} cm^{-3}$. A 1.5 nm SiO$_{2}$ cover is assumed. Next we discuss more in detail the procedure used to calculate $E_{b}$ and $S_{AA}$.

\subsection{Extraction of Barrier Height and the Active Cross Area Section} \label{sec_mod_Eb}

For pure thermionic emission any carrier energetic enough to surmount the barrier from the source to the channel (C) (Fig.~\ref{fig:Finfet_cartoon} (c)) will reach the drain provided the transport in the channel is close to ballistic. The Source/Drain in FETs are typically close to thermal and electrical equilibrium (since heavy scattering in the contacts is assumed which leads to instantaneous carrier relaxation). This allows us to make the assumption that most of the carriers in the Source/Drain  are thermalized at their respective Fermi-levels ($E_{fs}$, $E_{fd}$ in Fig. \ref{fig:Finfet_cartoon} (c)). Also the channel potential ($U_{scf}$) can be determined under the application of $V_{G}$ using the self-consistent scheme \cite{Paul1,Lee1}. Hence, for the source-to-channel homo-junction inside a FET, the barrier height ($E_{b}$) can be determined as a function of $V_{G}$,

\begin{equation}
\label{eqn_barht}
 E_{b}(V_{G}) = U_{scf}(V_{G}) - E_{fs}.
\end{equation}

\noindent This definition of $E_{b}$ implicitly contains the temperature dependence since the simulations are performed at different temperatures ($T$) which feature in the Fermi distribution of the Source/Drain, but, as it will be shown in section  \ref{sec_result}, the temperature dependence of $E_{b}$ in the sub-threshold region is very weak. Therefore, all the theoretical $E_{b}$ results shown in this section are at $T$ $\approx$ 300 K.

The study of thermionic emission model is applicable when the barrier height is much larger than the thermal broadening ($E_{b} \gg k_{B}T$ \cite{sze1}). For this reason, Eq. (\ref{eqn_barht}) works only in the sub-threshold region where $E_{b}$ is well defined \cite{Paul1} and once the FinFET is above the threshold, $E_{b}$ ($\le K_{B}T$) is not a well defined quantity anymore \cite{Paul1}. Furthermore, when the cross-section size of the FinFET is not large enough (i.e.: $W$, $H$ $\leqslant$ 20 nm) to be considered in a 3D bulk limit, $S_{AA}$ cannot be extracted using Eq. (\ref{Gbulk}) since the system is close to 1D. For a 1D system the $G$, under a small drain bias ($V_{SD}$) at a temperature T, is given by the following (for a single energy band),

\begin{equation}
\label{G1D}
   G_{1D} = \frac{2e^2}{h}\cdot\Big[1+exp(\frac{E_{b}(V_{G})}{k_{B}T})\Big]^{-1}
\end{equation}

\noindent where h is the Planck's constant. Since Eq. (\ref{G1D}) lacks any area description, G for 1D systems is no more a good method to extract $S_{AA}$. Below we will present an approach to solve this problem and to distinguish a 1D system from a 3D system. A part of all these limitations and as described in the previous sections \ref{sec:II} and \ref{sec:III}, $S_{AA}$ can be extracted using Eq. (\ref{Gbulk}).

\subsection{Trap extraction methods}
\label{extract_sec}

In Ref.~\cite{Tet150}, see also section \ref{sec:III}, it was observed that the active cross-section area ($S_{AA,sim}$) obtained theoretically is over-estimating the experimental value ($S_{AA,exp}$). In the results section \ref{sec_result} it will be further shown that also the theoretical $E_{b}$ value can over estimate the experimental $E_{b}$ value. These mismatches can be attributed to the presence of traps at the oxide-channel interface of multi-gate FETs where these traps can enhance the electro-static screening and suppress the action of the gate on the channel \cite{Kap232,Lee186,Tet150}. This simple idea is a powerful tool used for the estimation of interface trap density ($D_{it}$) in these undoped Si n-FinFETs.

\textbf{Method I: Interface Trap Density from Active Area}


\noindent Based on the difference between the simulated and the experimental active area ($S_{AA}$) values, a method to calculate the density of interface trap charges, $\sigma_{it}$, in the FinFETs is outlined. The method is based on the assumption that the total charge in the channel at a given $V_{G}$ must be the same in the experiments and in the simulations. This requirement leads to the following,

\begin{equation}
\label{charge_equality}
S_{AA,sim}\cdot L_{ch} \cdot \rho_{sim} = S_{AA,expt}\cdot L_{ch} \cdot \rho_{expt} + e\cdot \sigma_{it}\cdot L_{ch} \cdot P
\end{equation}

\noindent where $S_{AA,sim}$ ($S_{AA,expt}$) is the simulated (experimental) active area, $P$ is the perimeter of the channel under the gate ($P$ = $W$ + 2$H$) and $\rho_{sim}$ ($\rho_{expt}$) is the simulated (experimental) charge density. Close to the oxide channel interface it is possible to locally assume that $\rho_{expt}$ is obtained from $\rho_{sim}$ and $\sigma_{it}$ as,

\begin{equation}
 \label{rhoexp} 
 \rho_{expt} = \rho_{sim} - \rho_{it} = \rho_{sim} - (e \cdot \sigma_{it} \cdot P)/(W\cdot H)
\end{equation}

\noindent Using (\ref{charge_equality}) and (\ref{rhoexp}) the final expression for $\sigma_{it}$ is obtained as,

\begin{eqnarray}
 \label{trap_den1}
\sigma_{it}(V_{G}) & = & \frac{\rho_{sim}(V_{G})S_{AA,sim}(V_{G})}{e \cdot P} \\ \nonumber
		   & &\times \left[\frac{\left[1-\frac{S_{AA,expt}(V_{G})}{S_{AA,sim}(V_{G})}\right]}{\left[1-\frac{S_{AA,expt}(V_{gs})}{W\cdot H}\right]}\right] \;\; [\#/cm^2]
\end{eqnarray}

\noindent This method is useful for wider devices for which Eq. (\ref{Gbulk}) is valid. For very thin FinFETs (close to a 1D system) this method cannot be utilized .

\noindent \textit{Assumptions in Method I:} In the calculation of  $\sigma_{it}$ several assumptions were made. The extra charge contribution completely stems from the interface trap density ($D_{it}$) and any contribution from the bulk trap states has been neglected. Also all the interface traps are assumed to be completely filled which implies $\sigma_{it}$ $\cong$ $D_{it}$. This method of extraction works best for undoped channel since any filling of the impurity/dopant states is neglected in the calculation. Also the interface trap density is assumed to constant for the top and the side walls of the FinFET which is generally not the case \cite{Kap232,Lee186}. Orientation dependent $D_{it}$ for different surfaces could be included as a further refinement.

\textbf{Method II: Interface Trap Density from barrier control}


\begin{figure}[htb!]
\centering
\includegraphics[width=3in,height=2.2in]{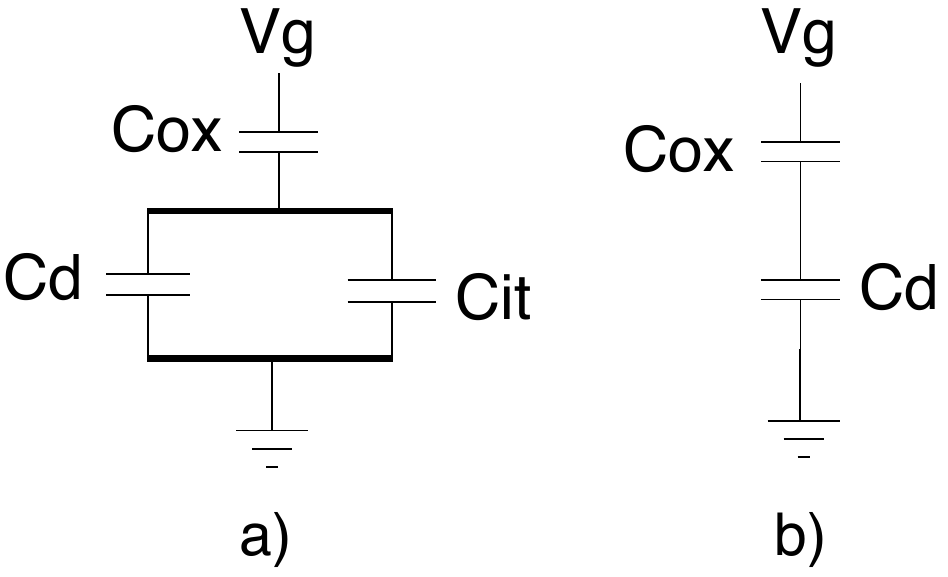}
\caption{Equivalent circuits (a) with interface-trap capacitance ($C_{it}$) and (b) without interface capacitance. $C_{d}$ and $C_{ox}$ are the depletion and the oxide capacitance, respectively. The idea for this equivalent circuit is obtained from page 381 in Ref.~\cite{sze1}. }
\label{C_trap_fig}
\end{figure}

\noindent The second method does not utilize the $E_{b}$ value directly but its derivative w.r.t $V_{G}$. The term $\alpha$ = $|dE_{b}/dV_{G}|$ represents the channel to gate coupling \cite{Sel073502,Tet150}. The presence of interface traps weakens this coupling due to the electrostatic screening. This method of trap extraction is based on the difference in the experimental and the simulated $\alpha$ value. The $\alpha$ value can be represented in terms of the channel and the oxide capacitance. The equivalent capacitance model for a MOSFET with and without interface traps ($D_{it}$) is shown in Fig. \ref{C_trap_fig}. The $\alpha$ value can be associated the oxide, interface and semiconductor capacitance which is given in Eq. (38) on page 383 in Ref. \cite{sze1}. This relation leads to the following,

\begin{equation}
\label{dEb_dEvg}
\vert \frac{dE_{b}}{dV_{G}} \vert = 1 - \frac{C_{tot}}{C_{ox}},
\end{equation} 

\noindent where $C_{tot}$ is the total capacitance. For the two cases, as shown in Fig.~\ref{C_trap_fig}, the total capacitance is given by, 

\begin{eqnarray}
\label{exp_c} C^{exp}_{tot} & = & \frac{C_{ox}\cdot (C_{d}+C_{it})}{C_{d}+C_{ox}+C_{it}}, \\
\label{sim_c} C^{sim}_{tot} & = & \frac{C_{d}\cdot C_{ox}}{C_{d}+C_{ox}},
\end{eqnarray} 

\noindent where $C_{it}$, $C_{ox}$ and $C_{d}$ are the interface trap capacitance, the oxide capacitance and the semi-conductor capacitance, respectively. Eq. (\ref{exp_c}) represents the capacitance in the experimental device and Eq. (\ref{sim_c}) represents the capacitance in the simulated device under ideal conditions without any interface traps. Combining Eq. (\ref{dEb_dEvg}), (\ref{exp_c}) and (\ref{sim_c}) and after some mathematical manipulations, it is possible to obtain,

\begin{equation}
\label{alpha_relation}
\frac{1}{\alpha_{exp}}  = \frac{1}{\alpha_{sim}} + \frac{C_{it}}{C_{ox}}, 
\end{equation}

\noindent Manipulating Eq. (\ref{alpha_relation}) gives the following relation for $C_{it}$,
\begin{equation}
\label{cit}
C_{it} =  C_{ox} \cdot \Big(\frac{1}{\alpha_{sim}}\Big)\cdot\Big[\frac{\alpha_{sim}}{\alpha_{exp}}-1\Big]
\end{equation}
Also $C_{it}$ can be related to the interface charge density ($\sigma_{it}$) as \cite{sze1},
\begin{equation}
\label{cit_2}
C_{it}  =  e \cdot \frac{\partial \sigma_{it}}{\partial V_{G}} 
\end{equation}

\noindent where $e$ is the electronic charge. In Eq. (\ref{cit}) all the values are dependent on $V_{G}$ except $C_{ox}$. Combining Eq. \ref{cit} and \ref{cit_2} and integrating w.r.t $V_{G}$ yields the final expression for the integrated interface charge density in these FinFETs as,

\begin{eqnarray}
\label{method2_eq}
\sigma_{it}& = & \frac{C_{ox}}{e}\cdot \int_{V1}^{V2=V_{th}} \Big(\frac{1}{\alpha_{sim}(V_{G})}\Big) \\ \nonumber
	   &   & \times \Big[\frac{\alpha_{sim}(V_{G})}{\alpha_{exp}(V_{G})}-1\Big] dV_{G}\;\; [\#/cm^{2}],
\end{eqnarray}

\noindent where $V_{th}$ is the threshold voltage of the FinFET and V1 is the minimum $V_{G}$ for which $\alpha_{exp/sim}$ is $\approx 1$. Of course, the integration range for Eq.~(\ref{method2_eq}) is in the sub-threshold region. This method has the advantage that it is independent of the dimensionality of the FinFET. Hence, Eq. (\ref{method2_eq}) can be used for wide as well as for thin FinFETs.

\noindent \textit{Assumptions in Method II:}  The  most important assumption is that the rate of change of the surface potential ($\Psi(V_{G})$) is the same as $E_{b}$ w.r.t $V_{G}$. The extra charge contribution completely originates from the density of interface trap charges ($\sigma_{it}$) and any contribution from the bulk trap states have been neglected. Also all the interface traps are assumed to be completely filled which implies $\sigma_{it}$ = $D_{it}$. This method works best when the change in the DC and the AC signal is low enough, such that the interface traps can follow the change in the bias sweep \cite{sze1}.

\textbf{Limitations of the methods}

\noindent To apply these trap metrology methods properly, is important to understand their limitations, which are presented in this section. One of the main limitation is how closely the simulated FinFET structure resembles the experimental device structure. This depends both on the SEM/TEM imaging as well the type of simulator used. In the present case a FinFET cross-section structure is created by using the TEM image making the simulated structure as close to the experimental device as possible. With the development of better TCAD tools, the proximity of the simulated structure to experimental structure has increased. This allows good confidence in the simulated conductance values then used for the interface trap calculations. Furthermore, the simulated $G$ is calculated as close to ideal as possible and all the differences between the ideal and experimental $G$ are attributed to the traps, which may not be true always. An important difference between the two methods is that they are calculated over different $V_{G}$ ranges. This is important since the trap filling and their behaviour changes within the $V_{G}$ range which should be taken into account accurately. One must also be aware of the embedded assumption of complete interface trap filling and the neglect of the bulk traps.

\subsection{Results and discussion}\label{sec_result}

In this section the theoretical results as well as their comparison with the experimental data are provided and discussed. 

\begin{table*}[hbt]
\centering
\begin{tabular}{|c|c|c|c|c|}\hline
Device	&	Method	&	$D_{it}\;(10^{11}cm^{-2})$	&	FET type		&	Obs. \\\hline

L=140nm*&	Charge	&	1.725					&	Special body  	& 	-- \\\cline{1-1} \cline{3-3} \cline{5-5}  
L=240nm*&	Pumping	&	2.072					&	tied FET		& 	-- \\\hline
A		&	I		&	5.560					&	Std. FET		&	$H_{2}$\\\cline{1-4}
\multirow{2}{*}{B} &	I 	&	10.60					&	Std. FET		&	anneal,  \\\cline{2-4}
		&	II		&	8.860					& Std. FET&reduces $D_{it}$ \\\hline
C 		   & II & 9.26 &  Std. FET & Thin fin, more $D_{it}$  \\\hline 

D                  & II & 18.31 & Std. FET& (110) side-wall, \\\cline{1-4}
\multirow{2}{*}{E} & I  & 18.1 & Std. FET &  thin fin, \\\cline{2-4}
 		   & II & 15.3 & Std. FET &  more etching, \\\cline{1-4}
F 		   & II & 36.3 & Std. FET &  more $D_{it}$ \\\hline
G  & II & 4.33 & Std. FET  & $H_{2}$ anneal, less $D_{it}$ \\\hline 
\end{tabular}
\caption{Values of $D_{it}$ obtained from all the n-FinFETs ($^{\ast}$ from Ref. \cite{Kap232}).}
\label{table2}
\end{table*}


\textbf{Temperature dependence of the Barrier Height}

\begin{figure}[hbt!]
	\centering
		\includegraphics[width=3.5 in,height=2.7 in]{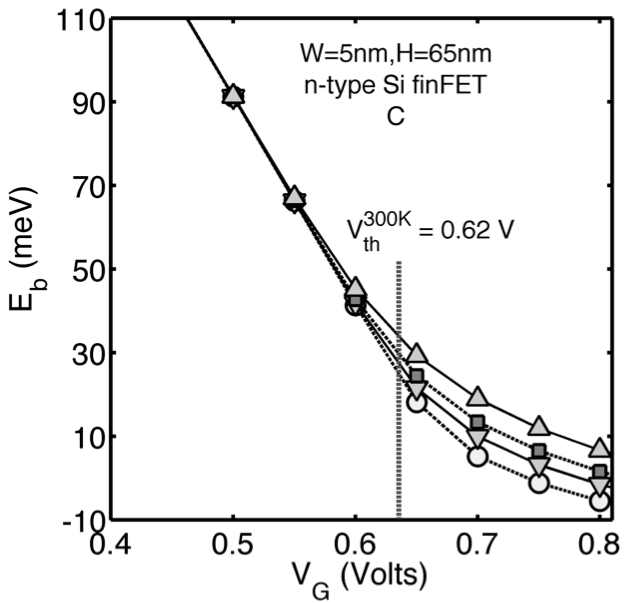}
	\caption{Temperature dependence of the simulated barrier height ($E_{b}$) in the n-FinFET C from 140 K to 300 K, (circle are for 140 K, down triangles for 200 K, squares for 240 K and up triangles for 300 K). At T=300K, $V_{th}$ of the FinFET is 0.62V. The overlap of the curves at different temperatures with $V_{G}$, below $V_{th}$ at 300K, shows a weak temperature dependence of $E_{b}$ in the sub-threshold region. The impact of temperature becomes prominent after $V_{G}$ goes above $V_{th}$.}
	\label{fig:Eb_dev_E_temp}
\end{figure}

\noindent The source-to-channel barrier height has been assumed to be temperature independent in the sub-threshold region. Figure \ref{fig:Eb_dev_E_temp} shows the results of a temperature dependent ToB calculations and proves that the barrier height ($E_{b}$) is only weakly temperature dependent in the sub-threshold regime. In the subthreshold region, the $E_{b}$ value for FinFET C, is same at four different temperatures ( $T$ = 140 K, 200 K, 240 K and 300 K). The variation with temperature becomes more prominent when the FinFET transitions into the on-state. Since, $E_{b}$ has a weak temperature dependence in the sub-threshold region it is then possible to evaluate $E_{b}$ from the 300 K simulations only.

\textbf{Evolution of the Barrier Height and of the Active Cross Section Area with $V_{G}$}  

\begin{figure*}[hbt!]
	\centering
		\includegraphics[width=5.6 in,height=3 in]{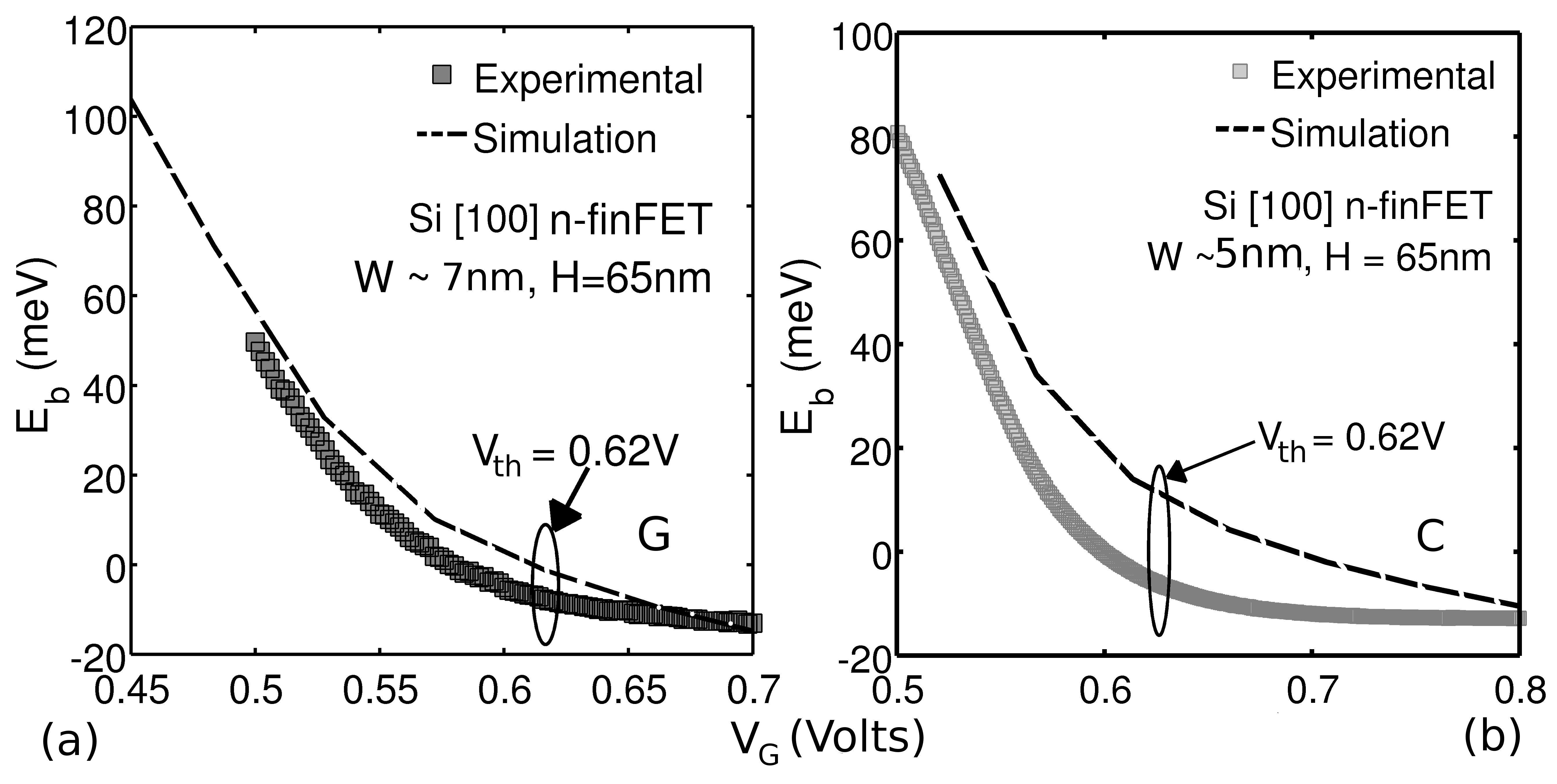}
	\caption{Experimental and simulated barrier height ($E_{b}$) in n-FinFET (a) G and (b) C. Both the devices have same $V_{th}$. Both experiment and 
	         simulation show a decreasing value of $E_{b}$ with $V_{G}$, but the absolute values are different. }
	\label{fig:Eb_dev_C_D}
\end{figure*}

\begin{figure*}[hbt!]
	\centering
		\includegraphics[width=5.6 in,height= 3 in]{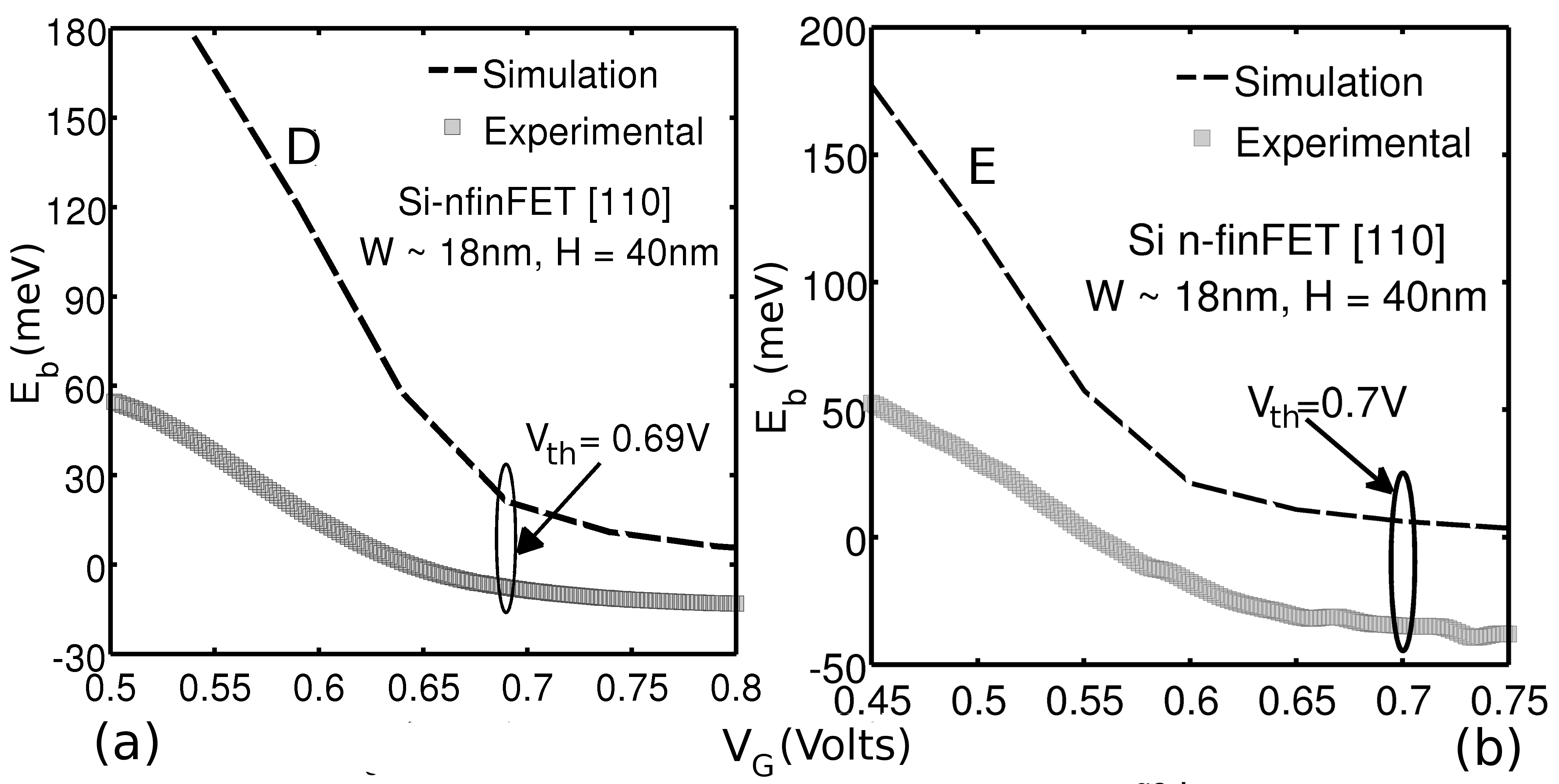}
	\caption{Experimental and simulated barrier height ($E_{b}$) in n-FinFETs (a) D and (b) E. Both the devices have similar $V_{T}$. Both experiment and 
	         simulation show a decreasing value of $E_{b}$ with $V_{G}$, but the absolute values are different. }
	\label{fig:Eb_dev_E_F}
\end{figure*}

\noindent Experimentally, it has been shown that, for undoped silicon n-FinFETs \cite{Tet150}, $E_{b}$ reduces as $V_{G}$ increases. Theoretically, the $E_{b}$ value is determined using Eq. (\ref{eqn_barht}) which depends on the self-consistent channel potential ($U_{scf}$). As the gate bias increases, the channel can support more charge. This is obtained by pushing the channel conduction band lower in energy to be populated more by the source and drain Fermi level \cite{Neo1286}. Figure~\ref{fig:Eb_dev_C_D} and \ref{fig:Eb_dev_E_F} show the experimental and theoretical evolution of $E_{b}$ in FinFETs G, C and D, E, respectively. Theory provides correct quantitative trend for $E_{b}$ with $V_{G}$. \textit{Few important observations here are, (i) theoretical $E_{b}$ value is always higher than experimental value and (ii) [110] Si devices (D and E) show larger mismatch to the experimental values.} The reason for the first point is suggested to be the presence of interface traps in the FinFETs which screen the gate from the channel \cite{Tet150}. The second observation can be understood by the fact that [110] channels with (110) sidewalls have more interface trap density due to the higher surface bond density \cite{sze1} and bad etching on the (110) sidewalls \cite{Kap232}.

\begin{figure*}[hbt!]
	\centering
		\includegraphics[width=5.6 in,height=3 in]{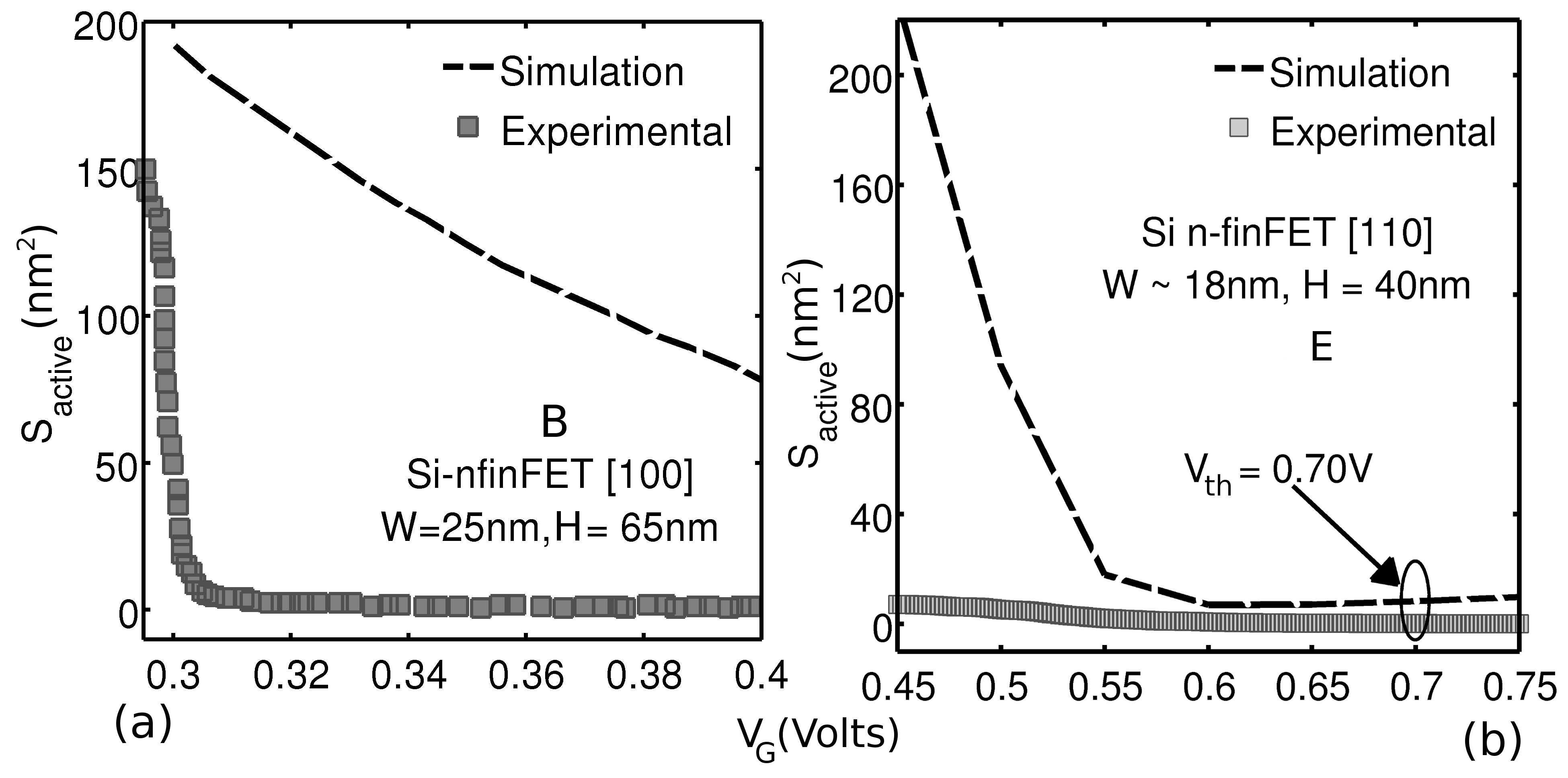}
	\caption{Experimental and simulated channel active cross-section ($S_{AA}$) in n-FinFETs (a) B and (b) E. Both experiment and 
	         simulation show a decreasing value of $S_{AA}$ with $V_{G}$, but the absolute values are different. }
	\label{fig:S_dev_B_F}
\end{figure*}

The active cross section area ($S_{AA}$) represents the part of the channel where the charge flows \cite{Tet150}. Experimentally $S_{AA}$ is shown to be decreasing with gate bias since the inversion charge moves closer to the interface which electrostatically screens the inner part of the channel from the gate \cite{Tet150,Tet2}. This gives a good indication of how much channel area is used for transporting the charge. Figure \ref{fig:S_dev_B_F} (a) and (b) show the experimental evolution of $S_{AA}$ in FinFET B and E, respectively. The theoretical value of $S_{AA}$ decreases with $V_{G}$ which is in qualitative agreement to the experimental observation \cite{Tet150}. However, the absolute values do not match. In fact theory over-estimates the experimental $S_{AA}$ value (Fig. \ref{fig:S_dev_B_F}) which is attributed to the interface traps.

\textbf{Trap density evaluation}

\begin{figure*}[hbt!]
	\centering
		\includegraphics[width=5.6 in,height=3 in]{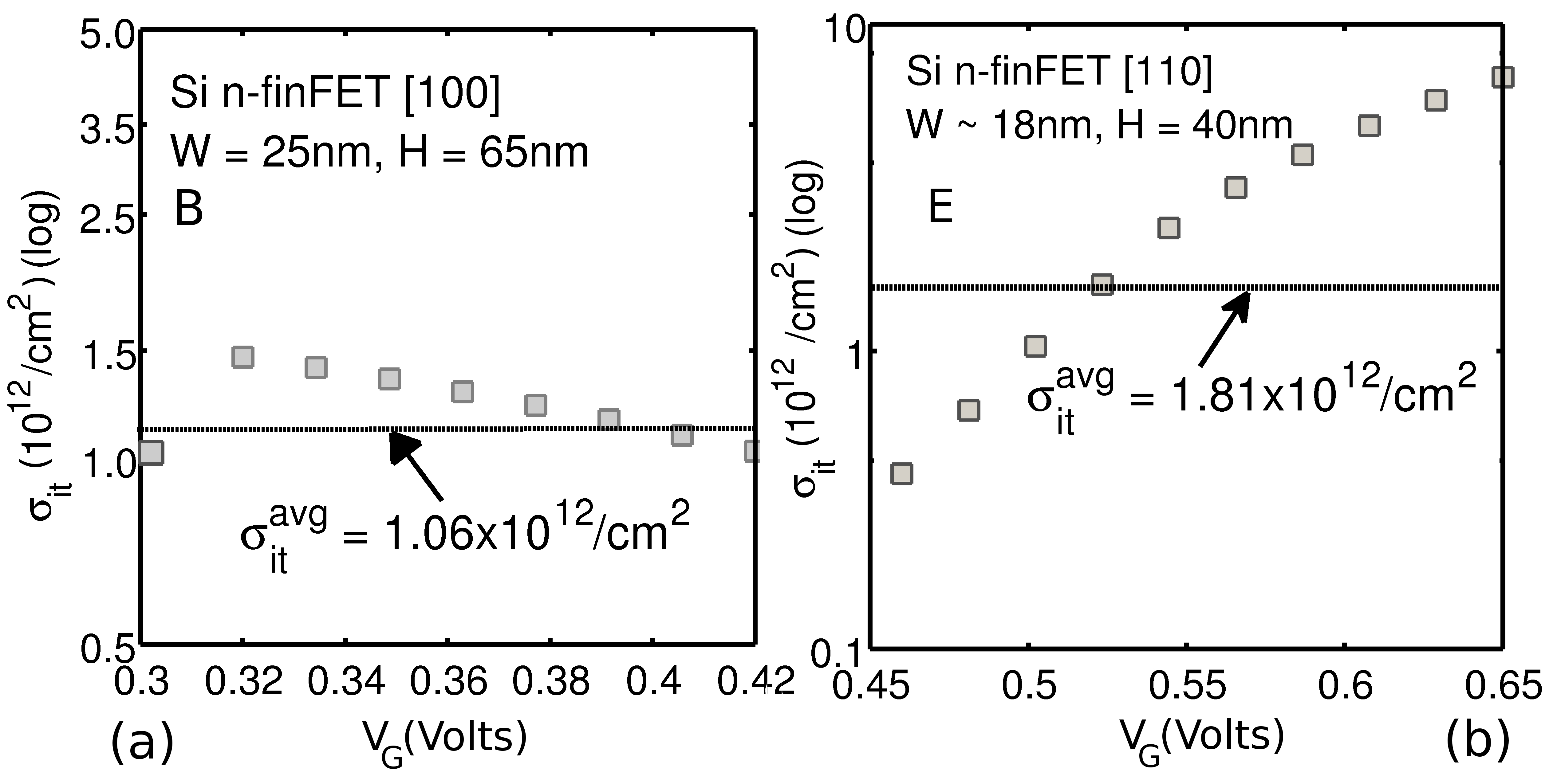}
	\caption{Extracted trap density using the difference in active device area (method I) for n-FinFETs (a) B and (b) E.}
	\label{fig:S_dit_B_F}
\end{figure*}

\begin{figure*}[hbt!]
	\centering
		\includegraphics[width=5.6 in,height=3 in]{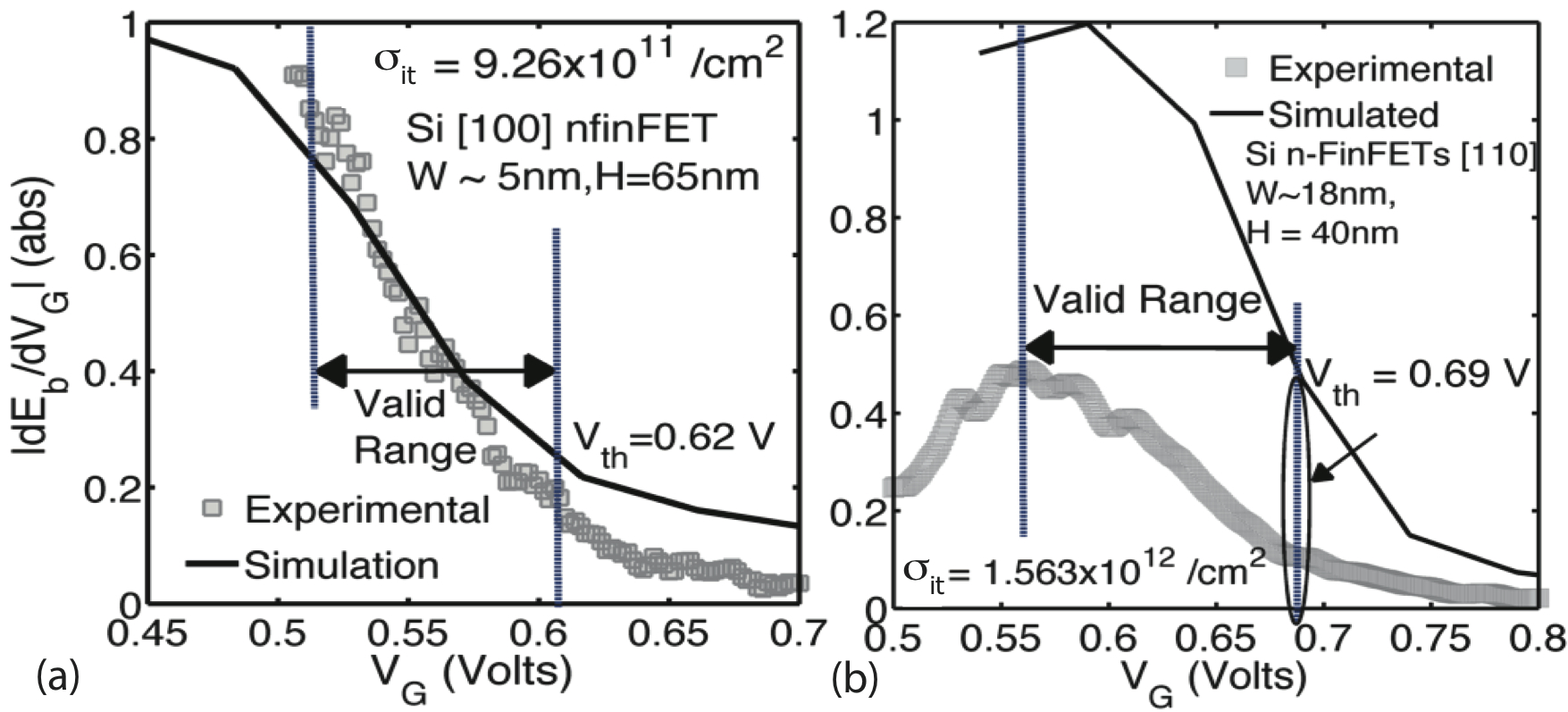}
	\caption{Experimental and simulated value of $\alpha$ in n-FinFETs (a) C and (b) E.}
	\label{fig:dEb_dvg_C_E}
\end{figure*}

\noindent In this section the results on $D_{it}$ in the undoped Si n-FinFETs are presented:


\textbf{$D_{it}$ using $S_{AA}$: $Method$ $I$}

\noindent This approach is based on method I (see section \ref{extract_sec} for details). The calculated $D_{it}$ values for FinFET B and E are 1.06e12$cm^{-2}$ and 1.81e12$cm^{-2}$ (Fig.\ref{fig:S_dit_B_F} (a) and (b), respectively). The $D_{it}$ values compare quite well with the experimental $D_{it}$ values presented in Ref. \cite{Kap232} and also shown in Table \ref{table2}. As expected the $D_{it}$ value for FinFET E (with [110] channel and (110) sidewalls) is higher than FinFET B ([100] channel with (100) sidewalls). This is attributed to the higher $D_{it}$ ($\sim 2\times$) on the (110) surfaces \cite{Kap232}. The results presented in this section show $\sim 1.8\times$ more $D_{it}$ for (110) sidewalls, in close agreement to previous experiments \cite{Kap232}. This method allows to calculate the $D_{it}$ in the actual FinFETs rather than custom made FETs.

\textbf{$D_{it}$ using $|dE_{b}/dV_{G}|: Method$ $II$}

\noindent This approach is based on method II (see section \ref{extract_sec} for details). The $C_{ox}$ value, needed in this method, is taken as $\sim$0.0173 $F/m^{2}$ which is assumed to be the same for all the devices since these FinFETs have similar oxide thickness. The calculated $D_{it}$ values for FinFET C and E are 9.26e11$cm^{-2}$ and 1.563e12$cm^{-2}$ (Fig.\ref{fig:dEb_dvg_C_E} (a) and (b), respectively). These calculations also show that [110] channel device (FinFET E) shows higher $D_{it}$ compared to the [100] channel device (FinFET C), again consistent with the observations made in Ref. \cite{Kap232}. The advantage of this method is that it can be used to obtain $D_{it}$ in extremely thin FinFETs (close to 1D system) unlike method I which is applicable only to wider FinFETs (due to the reasons discussed in section \ref{sec_mod_Eb}).


\textbf{Discussion of the two methods and $D_{it}$ trends}

\noindent{The $D_{it}$ values for all the FinFETs used in this study are shown in Table \ref{table2}. The important outcomes about the two methods are outlined below:}
\begin{itemize}
\item{The $D_{it}$ values obtained by the two methods compare very well with the experimental measurement in Ref. \cite{Kap232} for similar sized FinFETs (A and B). This shows the validity of these new methods.}
\item{The $D_{it}$ values calculated using method I and II (for B and E) compare very well with each other which shows that the two methods are complimentary \cite{Tet2}.}
\item{The $D_{it}$ values calculated for the two similar FinFETs (E and F) compare very well showing the reproducibility of the methods.}
\end{itemize}

\noindent The calculated $D_{it}$ values also reflect some important trends about the FinFET width scaling and surfaces (Table \ref{table2}). The central points are :

\begin{itemize}
\item{Hydrogen passivation considerably reduces $D_{it}$ \cite{Lee186}. This is observed for FinFETs A and B where $H_{2}$ passivation results in $\sim 2 \times$ less $D_{it}$ in FinFET A.}
\item{Width scaling requires more etching which also increases $D_{it}$ \cite{Kap232}. The same trend is observed in devices A to C and D to F (decreasing $W$).}
\item{(110) sidewalls show higher $D_{it}$ compared to (100) sidewalls \cite{Kap232}. The same trend is also observed for FinFETs A, B, C, G ((100) sidewall) compared to FinFETs D, E and F ((110) sidewall).} 
\end{itemize}

\subsection{\textbf{Current distribution}}\label{sec_current_res}

\begin{figure*}[htb]
	\centering
		\includegraphics[width=4. in,height=2.6 in]{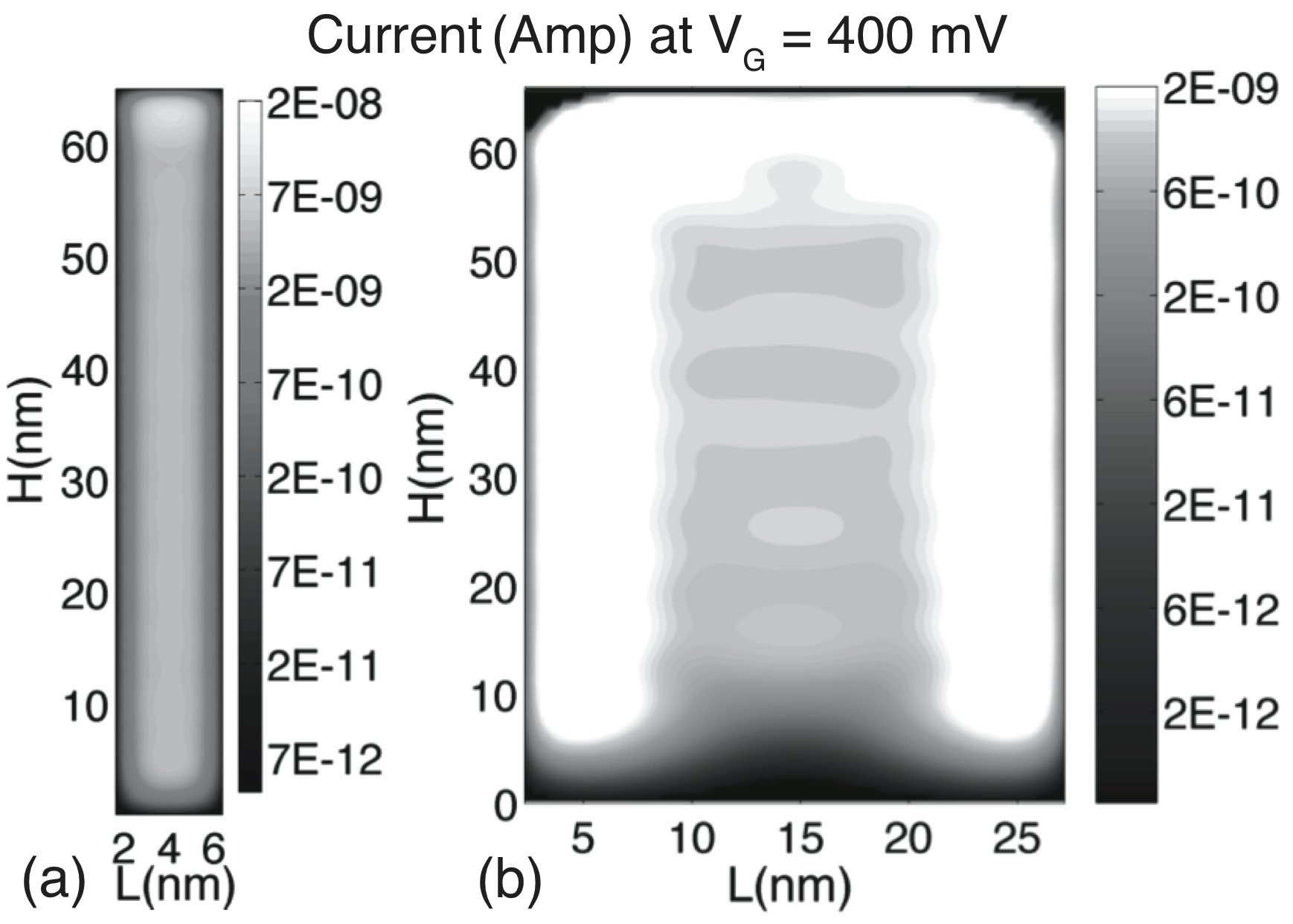}
	\caption{Simulated spatial current distribution in the  [100] undoped Si n-FinFET intrinsic with H = 65nm and (a) W = 5nm and (b) W = 25nm. $V_{G}$ = 0.4V and $V_{SD}$ = 30mV at 300K. 5nm device shows a complete volume inversion. In the 25nm device the current mainly flows at the edges.}
	\label{fig:curr_dist}
\end{figure*}

The charge flow in n-FinFETs show a very strong dependence on the geometrical confinement. For very small width FinFET the entire body gets inverted and shows a very little change in $S_{AA}$ with $V_{G}$. For wider FinFETs the current flow starts from a weak volume inversion and moves to surface inversion as $V_{G}$ increases \cite{Tet150}. The theoretical spatial current calculation reveals similar trend which is shown in Fig. \ref{fig:curr_dist} For extremely thin n-FinFETs ($W$ = 5 nm, $H$ = 65 nm) the charge flow is prevalently through the entire body (volume inversion) compared to the wider n-FinFETs ($W$ = 25 nm, $H$ = 65 nm) where the charge flows at the edges. This reflects the fact that thinner FinFETs show better channel area utilisation for the charge flow. However, an important practical limitation comes from the fact that extremely thin FinFETs also require more etching, which increases $D_{it}$ and hence can limit the action of thin FinFETs. The advancement of fabrication methods and strain technology may improve the performance of thin FinFETs as shown by some experimental works \cite{Won133}. 

\subsection{Conclusion of section IV}\label{sec_conc}

A new $D_{it}$ determination methodology for state-of-the-art n-FinFETs is presented. Two complementary approaches provide (a) the gate bias ($V_{G}$) dependence of $D_{it}$ (Method I) and, (b) the total $D_{it}$ (Method II).

\noindent The following trends are observed:

\begin{itemize}
\item (i) The hydrogen annealing step in the fabrication process substantially reduces $D_{it}$ in good agreement with Ref. \cite{Lee186}
\item (ii) The scaling of the $W$ of the devices (i.e.: from $A$ to $C$ or from $D$ ($E$) to $F$) increases the density of  interface states
\item (iii) The change in the orientation of the channel (and therefore the sidewall surface where the interface traps are formed) from $[$100$]$ (device $A$ or $C$)  to $[$110$]$ (device $D$ ($E$) or $F$) remarkably increases the density of interface states
\item (iv) By comparison of the value of $D_{it}$ obtained for device $B$ in the two approaches (i.e.: see Fig. \ref{fig:S_dev_B_F} and Table \ref{table2}) and the value of $D_{it}$ obtained for two identical devices ($D$ and $E$) using the same approach (Method II), compatibility and reproducibility of the methods are demonstrated. 
\end{itemize}

The reported trends are similar to the one suggested in the literature \cite{Lee186,Xio541}. The simple Top-of-the-barrier model, combined with Tight-binding calculations, explains very well the thermally activated sub-threshold transport in state-of-the-art Si FinFETs. The qualitative evolution of $E_{b}$ and $S_{AA}$ with $V_{G}$ are well explained by the theory. Furthermore, the mismatch in the quantitative values of $E_{b}$ and $S_{AA}$ led to the development of two new interface trap density calculation methods. The advantage of these methods is that they do not require any special structure as needed by the present experimental methods. Hence the interface quality of the ultimate channel can be obtained. These methods are shown to provide consistent and reproducible results which compare very well with the independent experimental trap measurement results. The calculated trends of interface trap density with channel width scaling, channel orientation and hydrogen passivation of the surfaces compare well with the experimental observations. The volume inversion observed in thin width FinFETs is more efficient, in term of volume utilisation. However, it could lead to a better utilisation of FETs channel only if surfaces roughness and the density of interface traps, created during the extreme etching necessary for these device to be fabricated, can be reduced. 

\section{Final conclusions} \label{final}
Overall, this paper discusses how, by making use of a classical tool such as the thermionic emission theory in combination with state-of-the-art tight binding simulations, it is possible to provide precious information on the transport characteristics of ultra-scaled Si n-FinFETs. In fact, it is demonstrated here that the amplitude of the energy barrier, of the region of transport in the channel and of the interface trap density, are all quantities that can be directly estimated in state-of-the-art FinFETs. Due to the rapid scaling of CMOS-FET technology, the techniques introduced in this paper could become routine tools for device improvement and optimisation.

\section{Acknowledgments}

G.C.T. and S.R. acknowledge financial support from the EC FP7 FET-proactive NanoICT projects MOLOC (215750) and AFSiD (214989), the Dutch Fundamenteel Onderzoek der Materie FOM. G.C.T., J.V., L.C.L.H. and the ARC-CQC2T (project number CE110001027). A.P., S. L. and G. K. acknowledge financial support from U.S. National Security Agency (NSA) and the Army Research Office (ARO) under Contract No. W911NF-04-1-0290. The work at Purdue and JPL is supported by grants from the Army Research Office. The work at the Jet Propulsion Laboratory, California Institute of Technology, is supported by grants from the National Aeronautics and Space Administration. 
\newpage

\bibliography{bib}

\end{document}